\documentclass[review]{elsarticle}

\let\today\relax
\makeatletter
\def\ps@pprintTitle{%
    \let\@oddhead\@empty
    \let\@evenhead\@empty
    \def\@oddfoot{\footnotesize\itshape
         {Submitted preprint} \hfill\today}%
    \let\@evenfoot\@oddfoot
    }
\makeatother

\usepackage{lineno}
\usepackage{hyperref}
\usepackage{enumerate}
\usepackage{amssymb}
\usepackage[italicdiff]{physics}
\usepackage{graphicx}
\usepackage{subfigure}
\usepackage{float}
\usepackage{courier}

\usepackage{listings}
\usepackage{color}
\definecolor{dkgreen}{rgb}{0,0.6,0}
\definecolor{gray}{rgb}{0.5,0.5,0.5}
\definecolor{mauve}{rgb}{0.58,0,0.82}
\usepackage{array, makecell}

\begin{document}

\begin{frontmatter}

\title{Multi-GPU Performance Optimization of a CFD Code using OpenACC on Different Platforms}

\author{Weicheng Xue\corref{cor1}\fnref{fn1}}
\ead{weich97@vt.edu}
\author{Christoper J. Roy\fnref{fn3}}
\ead{cjroy@vt.edu}

\cortext[cor1]{Corresponding author}
\fntext[fn1]{Graduate Assistant, AIAA Student Member}
\fntext[fn3]{AIAA Associate Fellow}
\address{Virginia Tech Kevin T. Crofton Department of Aerospace and Ocean Engineering,\\
215 Randolph Hall, Blacksburg, VA 24061, US}

\begin{abstract}
This paper investigates the multi-GPU performance of a 3D buoyancy driven cavity solver using MPI and OpenACC directives on different platforms. The paper shows that decomposing the total problem in different dimensions affects the strong scaling performance significantly for the GPU. Without proper performance optimizations, it is shown that 1D domain decomposition scales poorly on multiple GPUs due to the noncontiguous memory access. The performance using whatever decompositions can be benefited from a series of performance optimizations in the paper. Since the buoyancy driven cavity code is latency-bounded on the clusters examined, a series of optimizations both agnostic and tailored to the platforms are designed to reduce the latency cost and improve memory throughput between hosts and devices efficiently. First, the parallel message packing/unpacking strategy developed for noncontiguous data movement between hosts and devices improves the overall performance by about a factor of 2. Second, transferring different data based on the stencil sizes for different variables further reduces the communication overhead. These two optimizations are general enough to be beneficial to stencil computations having ghost changes on all of the clusters tested. Third, GPUDirect is used to improve the communication on clusters which have the hardware and software support for direct communication between GPUs without staging CPU's memory. Finally, overlapping the communication and computations is shown to be not efficient on multi-GPUs if only using MPI or MPI+OpenACC. Although we believe our implementation has revealed enough overlap, the actual running does not utilize the overlap well due to a lack of asynchronous progression.
\end{abstract}

\begin{keyword}
Multi-GPU, OpenACC, MPI, Domain Decomposition, Performance Optimization, GPUDirect
\end{keyword}

\newpageafter{abstract}

\end{frontmatter}


\section{Introduction}

Computational Fluid Dynamics (CFD) is a method which can be used to solve physical problems in the field of fluids, usually requiring a lot of computation. In order to obtain more accurate numerical solutions for challenging problems, researchers are using very fine meshes or high-order schemes which require much better resources such as a larger memory and faster processor. However, generally the memory cannot be infinitely large and a processor cannot hold an infinite number of transistors. These limitations may require codes to be written in a data parallel way, i.e., decomposing a big problem into small pieces and distributing these small problems to multi/many-cores or even accelerators such as GPUs. Applying high performance computing (HPC) in the CFD~\cite{Using_HPC} area is necessary and has aroused CFD researchers' interest.

On multicore/manycore CPU systems or equivalent, there are three common paradigms for HPC: OpenMP, MPI, and hybrid MPI+OpenMP. OpenMP~\cite{OpenMP} is designed for shared memory systems so that data can be shared among all threads but this comes with the risk of race conditions to exist if multiple threads are modifying the same data. Also, scaling performance across multiple nodes or sockets may be poor on distributed memory systems (more usually used than shared memory systems). MPI~\cite{MPI} is a message passing standard designed for various platforms including shared and distributed memory architectures. Data is moved between processors through sending and receiving messages. However, programming with MPI is more complicated than with OpenMP as it requires extra care to decompose the problem well and implement efficient communications. Usually, communication may be a significant bottleneck when the codes is scaled up to a lot of processors. Hybrid MPI+OpenMP methods~\cite{MPI_OpenMP} are therefore proposed to combine the advantages of MPI and OpenMP. The hybrid model is a good match with modern multicore/manycore architectures, so it can be programmed efficiently using two levels of parallelism: MPI for the inter-node/socket communication and OpenMP for the intra-node/socket computation and communication. However, hybrid MPI+OpenMP cannot be easily used on GPUs directly due to a lack of full support of OpenMP 4.0 (or later) from the compiler development. 

In addition to the CPU computing, GPU computing have been gaining a lot of interests~\cite{GPUs_SC}. This attention is because of the GPU's high compute capability and memory bandwidth as well as their low power consumption. A single GPU have thousands of threads, therefore numerous threads can execute a same instruction simultaneously on multiple data points, known as single instruction multiple threads (SIMT). When executing a program on the GPU, the highly compute-expensive portion of a program is offloaded to the GPU. Then numerous threads on the GPU execute the code simultaneously to achieve a high speedup. File I/O, branch controls, printout, etc. remain on the CPU since the CPU has the flexibility to perform these tasks while GPUs are not as efficient for these complex tasks (if they are even possible). One important thing to mention here is that although GPUs provides higher memory bandwidth than CPUs, different memory access patterns may significantly affect the actual memory throughput~\cite{Memory_Access}, which should be considered carefully.

Three language extensions/libraries are widely applied to port codes to GPUs~\cite{memeti2017benchmarking}. They are OpenCL, CUDA and OpenACC. OpenCL and CUDA are C/C++ with some extensions while OpenACC is a compiler directive-based interface, similar to OpenMP. OpenCL and CUDA are low level programming models so that they require users to have some background in computer architecture systems. Also, programming with OpenCL or CUDA is difficult, as users need to rewrite and tune their codes on various GPUs every time. CUDA gets a strong support from NVIDIA but it cannot be compatible well with other GPUs such as AMD. OpenCL is open source and we found that it has not been commonly used in real world GPU-accelerated CFD codes, possibly due to the high complexity of programming and a lack of good ecosystem support. Different from OpenCL and CUDA, OpenACC is a high level programming model that enables users to accelerate their CFD codes more readily on various GPUs without intruding their legacy codes completely. Programmers using OpenACC~\cite{OpenACC} can be somewhat agnostic about the GPU architecture compared to using OpenCL and CUDA because compilers such as those developed by PGI (acquired by NVIDIA) can hide a lot of details and decide how to optimize the code (although it may not be optimal). Also, because of its directive-based feature and good support for portability, OpenACC can be much easier to use on various platforms compared to CUDA. We will also show this benefit because there is little code modification across platforms. However, to gain good performance across different platforms, the features of the architecture and some low level optimizations should be taken into account. Apart from the options mentioned, OpenMP can be a potential viable choice for the GPU once the development of compilers catch up in the future. Because OpenACC provides an easy way of programming~\cite{wienke2012openacc}, a good feature for portability across platforms~\cite{sabne2014evaluating} and good parallel performance if optimized enough~\cite{hoshino2013cuda}, OpenACC was applied to port our CFD code to the GPU.

OpenACC has already been used for various GPU-accelerated CFD codes or related applications. Gong et al.~\cite{gong2016nekbone} presented an optimized OpenACC version for NekBone, which is a core kernel of the incompressible Navier-Stokes solver Nek500, based on their group's prior work. They ported the optimized code to multiple GPU systems and obtained a parallel efficiency of 79.9\% on 1024 GPUs. However, the code they worked on is just a kernel, not a complete CFD code. Hoshino et al.~\cite{hoshino2013cuda} found that although OpenACC is 50\% slower than CUDA for a naive implementation, the gap can be decreased to only 2\% after careful manual optimizations. They also pointed out that there are some intrinsic deficiencies of OpenACC, such as a lack of interface for shared memory and inter-thread communication. Searles et al.~\cite{searles2018mpi+} studied a wavefront based mini-application for a production code for nuclear reactor modeling. It is interesting and rare to see in their work that the OpenACC implementation is even slightly faster than CUDA. Their work mainly focused on exploring complex parallel patterns in their code and exploring the scalability using MPI across different platforms. In summary, OpenACC is easy to use and also good for portability across different platforms, however to obtain good performance, careful pertinent optimizations for an application may need to be designed.

To assess the performance of a code accelerated by the CPU or GPU, weak scaling and strong scaling performance are often measured. The major difference between the two scalings is whether one keeps the total problem size fixed (strong scaling) or the problem size per processor fixed (weak scaling), while adding more processors. Obviously maintaining strong scaling is more challenging. Commonly both scalings are investigated to satisfy different situations such as solving a fixed problem as fast as possible (strong scaling) or solving as big of a problem as possible (weak scaling). In both situations we want to max out all compute nodes or resources to gain the maximum speedup. In the CFD area, we are more interested in the weak scaling performance as we hope to solve a larger and complex problem faster, if more compute resources are available. However, there are also a lot of occasions in which small problems need to be solved if the requirement for numerical accuracy is not high. Therefore, both the weak scaling and strong scaling performance are measured in this paper.

Prior to the work presented in this paper, Pickering et al.~\cite{pickering2015directive} examined the process of applying OpenACC to a 2D CFD code using both single precision and double precision. They also applied OpenMP's fork/join execution model to scale the performance up to 4 NVIDIA C2070 GPUs with a strong scaling efficiency of 95\%. Instead of using the OpenMP+OpenACC model, Baghapour et al.~\cite{baghapour2016multilevel} switched to the MPI+OpenACC model and scaled a 3D CFD code well up to both 32 CPUs and 32 GPUs on a distributed cluster. In their work, they used 1D domain decomposition to distribute the load to different processors and increased the grid size in only one dimension for their weak scaling performance. Xue et al.~\cite{xue2018multi} compared multi-CPU/GPU performance using 3D, 2D and 1D decompositions and gave a primitive analysis of their differences. Also, two performance optimizations including an pack/unpack method for data exchange between hosts and devices was designed, and this pack/unpack method was proved to improve the performance using 3D decomposition greatly on a platform using old GPUs (NVIDIA C2070). In the current paper, the effects of multiple factors such as the platform architecture, decomposition methods, pack/unpack, strong v.s. weak scaling and GPUDirect will be investigated. Also, the limitations of overlapping communication and computation if applying MPI or MPI+OpenACC are presented.

\section{CFD Code: Buoyancy Driven Cavity Solver}
The 3D Buoyancy Driven Cavity (BDC) problem has a cubic domain, a vertical wall and its opposing wall have different temperatures, and the horizontal walls are adiabatic. A gravitational force is added to the air in the square cavity. Heat flux caused by the temperature difference leads small density changes in the fluid (Boussinesq approximation), and the buoyancy effect (density change) causes the fluid to convect in the cavity.

The CFD code written with Fortran 2003/2008 in the paper solves the classic 3D BDC problem~\cite{baghapour2016multilevel}, which is a system of 3D incompressible Navier-Stokes equations. An artificial compressibility method developed by Chorin~\cite{chorin1997numerical} is used. The artificial viscosity term makes the system of equations to be hyperbolic so that steady state solution can be obtained through time marching. The CFD code use a first order Euler explicit scheme for temporal discretization, and a second-order central-difference scheme with an artificial dissipation term for spatial discretization. The artificial dissipation term is applied to the continuity equation to alleviate odd-even decoupling. The numerical damping term is based on the fourth-derivative of pressure and is discretized using a second-order central finite difference scheme. The discretized form of the system of the governing equations can be written as,
\begin{equation}
\label{governing}
\frac{1}{\beta^2}\frac{\partial p}{\partial t} +\rho\frac{\partial V_i}{\partial x_i}={\epsilon_i}\pdv[4]{p}{x_i}
\end{equation}
\begin{equation}
\label{governing2}
\frac{\partial {V_j}}{\partial t} +{V_i}\frac{\partial V_j}{\partial x_i}=-\frac{1}{\rho}\frac{\partial p}{\partial x_j}+\nu\pdv[2]{V_j}{x_i}+\sigma(T-T_{\infty}){g_j}
\end{equation}
\begin{equation}
\label{governing3}
\frac{\partial T}{\partial t} +{V_i}\frac{\partial T_j}{\partial x_i}=\alpha\pdv[2]{T_j}{x_i}
\end{equation}
where ${\beta}$ is an artificial compressibility parameter calculated using the local velocity magnitude along with a reference velocity defined by the user, ${\epsilon}$ are numerical dissipation coefficients controlling stability, ${\nu}$ is the kinematic viscosity of the fluid, and ${\alpha}$ is the thermal dissipation rate.

In this paper, the size of the cavity is 0.05 m in all three dimensions. Pressure is extrapolated to the ghost cells at adiabatic walls using a one-sided second order scheme. Temperature is similarly extrapolated to the horizontal wall ghost cells using a second order scheme. Pressure is rescaled at the center point of the cavity in every iteration. The meshes used for the BDC code are uniform and their size range from $32^3$ to $1024^3$. The Rayleigh number is set to be 100,000 for the convection problem. Most of the constant settings affecting the flow do not affect the parallel performance.

\section{Implementation}

\subsection{Stencil Computation}
\begin{figure}[H]
	\centering
	\includegraphics[width=.48\textwidth]{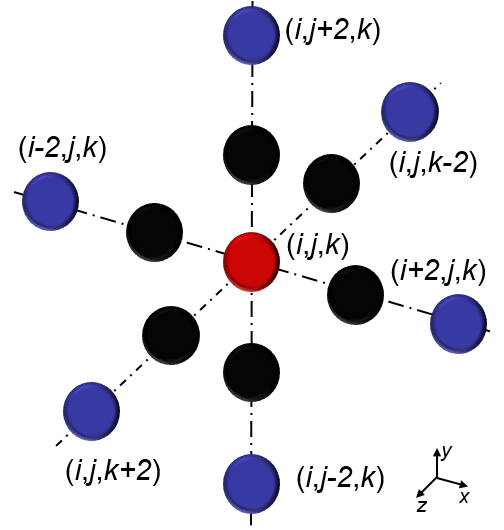}
	\caption{Stencil (black+red: velocity and temperature stencil, blue+black+red: pressure stencil)}
	\label{Stencil}
\end{figure}

In the BDC code studied here, since we use the fourth-derivative pressure dissipation term and a second-order central-difference scheme for all spatial derivatives, the numerical stencil size for the pressure is 5 and for other primitive variables is 3. This can be utilized to design an optimization to reduce the data exchange across processors (\emph{Optimized V2}, will be introduced later). Fig.~\ref{Stencil} shows the stencils for a node in the computational domain. The iterative residual calculation in the explicit CFD code is intrinsically one kind of stencil computation. For each node, it needs the data in its two stencils to compute and fill in the residual array, which is later used to update the primitive variables. There should be a nested loop over all nodes in the spatial domain, and a time step loop containing all stencil computations to iterate in pseudo time domain. When programming, data locality should be considered to use cache more efficiently. For a 3D array $A(i,j,k)$ in Fortran, $i$ is set to be incremented fastest (the innermost loop) and $k$ the slowest (the outermost loop). This layout is also good for the GPU, as the GPU prefers the coalesced memory access pattern in which contiguous threads in a thread block operate on the consecutive memory locations. It should be noted here that the index directions ($i$, $j$ and $k$) are aligned with the spatial directions ($x$, $y$ and $z$) in this problem. Therefore, the use of $(i, j, k)$ are mixed with the use of $(x, y, z)$ in the paper.

\subsection{Domain Decomposition}
\begin{figure}[H]
	\centering
	\includegraphics[width=.48\textwidth]{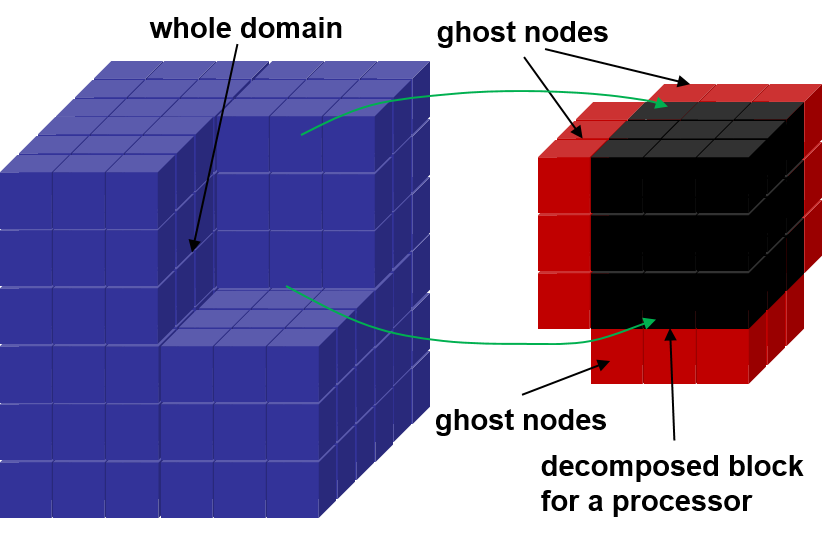}
	\caption{3D domain decomposition}
	\label{Decomposition}
\end{figure}

Many methods can be used to decompose a computational domain such as structured partitioning~\cite{ytterstrom1997tool} and graph partitioning \cite{rantakokko2000partitioning}. For a CFD problem with single-block structured grid such as a BDC problem running on pure CPUs or pure GPUs, there are three structured ways: 1D decomposition, 2D decomposition and 3D decomposition. Which way of decomposing the domain performs the best greatly depends on the application and computer architecture. On one hand, the surface-area/volume ratio determines the total size of data transferred between processors and the total size of ghost cells, and 3D decomposition has the lowest area/volume ratio (least ghost cells). On the other hand, the frequency of data transfers between processors can greatly affect the performance, especially when the memory bandwidth or latency issue becomes important, and 1D decomposition has the least times of data transfers. Besides, for stencil computations like ours, 1D or even 2D decomposition may generate too thin slices that invalidates the spatial discretization scheme if scaling up to a large number of processors (strong scaling). Thus, it is worthwhile to investigate the effect of various domain decompositions on different platforms. Xue et al.~\cite{xue2018multi} showed that 2D decomposition scales better up to 32 CPUs compared to 1D and 3D decomposition on a platform, and 3D decomposition can outperform 1D decomposition if applying optimizations. However, they did not show the comparison between 3D decomposition and 2D decomposition. Also, they only tested the code on a single platform having old GPUs. An example 3D decomposition adopted in this paper is shown in Fig.~\ref{Decomposition}. Each processor is given a decomposed block, with each cutting face contacting a neighbour block. Ghost nodes are used to store decomposed boundary information transferred from neighbours. 1D and 2D decompositions are similar as 3D decomposition but have fewer decomposed directions and fewer decomposed boundaries.

For a given number of processors, there may be many combinations for either 1D, 2D or 3D decomposition. For general situations, we try to decompose the domain evenly in all the available dimensions but decompose more in the slowest stride index direction, to preserve more contiguous data after decomposition. This method is designed to 1) divide the domain in the available dimensions as evenly as possible, 2) utilize enough processors, 3) decompose with a priority along the $k$ dimension first, then $j$, and finally $i$, as Fortran is a column-majored language.

\subsection{Hardware Configuration}
\paragraph{HokieSpeed}
Although now decommissioned in June 2017, HokieSpeed~\cite{Hokiespeed} was a cluster at Virginia Tech and was previously in the list of Green500. HokieSpeed~\cite{Hokiespeed} had 204 nodes using a quad data rate InfiniBand interconnect. Each node was outfitted with 24GB memory, two six-core Xeon E5645 CPUs and two NVIDIA M2015/C2050 GPUs. Every GPU had 14 multiprocessors (MP) and 3GB memory. The peak bandwidth to the 3GB shared memory was 148.4GB/s. Every MP had 32 CUDA cores, 48KB shared memory and 16KB L1 cache. All the access to the global memory went through the L2 cache of size 512KB. The peak double precision performance was 513 GFLOPS. The compilers used on HokieSpeed were PGI 15.7 and Open MPI 1.10.0. A compiler optimization of -O4 was used.

\paragraph{NewRiver}
NewRiver~\cite{Newriver} is a cluster at Virginia Tech (VT). It has 39 GPU nodes shared by the whole VT community. On NewRiver~\cite{Newriver}, each of these nodes is equipped with two Intel Xeon E5-2680v4 (Broadwell) 2.4GHz CPU (28 cores/node in all), 512 GB memory, and two NVIDIA P100 GPUs. Each NVIDIA P100 GPU is capable of up to a theoretical 4.7 TeraFLOPS of double-precision performance. The NVIDIA P100 GPU offers much higher GFLOPS and memory bandwidth compared with the NVIDIA C2050 GPU on HokieSpeed. The modules used on NewRiver are PGI 17.5, CUDA 8.0.61 and Open MPI 2.0.0 or MVAPICH2-GDR 2.3b. It should be mentioned that MVAPICH2-GDR 2.3b is a CUDA-aware MPI wrapper compiler which supports GPUDirect (if this feature is turned on). An compiler optimization of -O4 is used. The maximum number of nodes which can be used is 12, but only 8 is used in this paper. Thus, the maximum number of GPUs used on the NewRiver cluster is 16 (when using both Open MPI 2.0.0 or MVAPICH2-GDR 2.3b). If not specified, Open MPI 2.0.0 is used.

\paragraph{Cascades}
Cascades~\cite{Cascades} is another cluster at VT. It has 40 GPU nodes shared by the whole VT community. On Cascades~\cite{Cascades}, each of these nodes is equipped with two Intel Skylake Xeon Gold 3 Ghz CPUs (24 cores/node in all), 768 GB memory, and two NVIDIA V100 GPUs. Each NVIDIA V100 GPU is capable of up to 7.8 TeraFLOPS of double precision performance, which is 66\% higher than the P100 GPU on the NewRiver cluster. The NVIDIA V100 GPU offers the highest GFLOPS and memory bandwidth among the GPUs we used. The modules used on Cascades are PGI 18.1, CUDA 8.0.61 and Open MPI 3.0.0 or MVAPICH2-GDR 2.3b. An compiler optimization of -O4 is used. Similarly, the maximum number of GPUs used on the Cascades cluster is 16 when using Open MPI 3.0.0. However, when switching to MVAPICH2-GDR 2.3b, since Cascades uses "srun" to run MPI programs instead of using "mpirun\_rsh" (as Slurm is used on the Cascades), the maximum number of GPUs which can be used is only 8 (using more would cause the efficiency to drop to about 1\%, which is not reasonable and caused by some unknown issues related to Slurm). If not specified, Open MPI 3.0.0 is used.

\section{Results}\label{Results}

\subsection{BDC Solution}
\begin{figure}[H]
	\centering 
	\subfigure[3D pressure contour]{ 
		\label{3D_Pressure}
		\includegraphics[width=.45\textwidth]{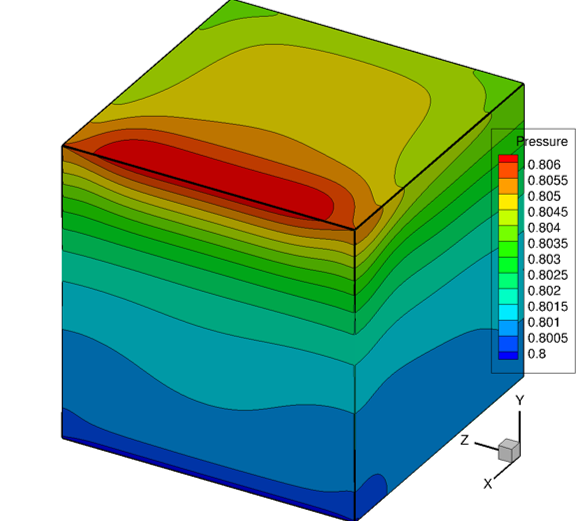} 
	} 
	\subfigure[2D temperature contour at y=0.025 m]{ 
		\label{2D_Temperature}
		\includegraphics[width=.45\textwidth]{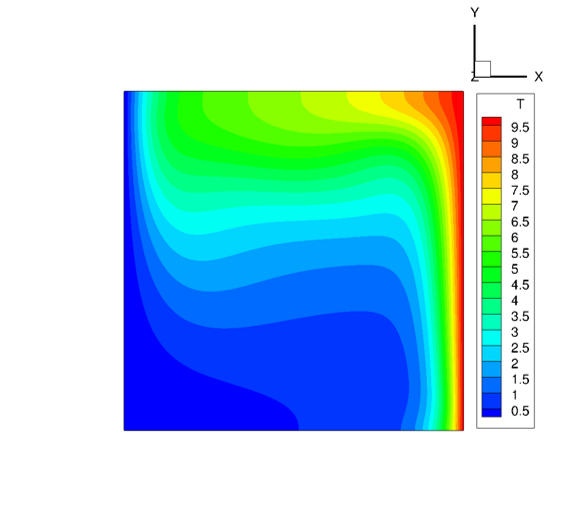} 
	} 
	\caption{3D BDC solution} 
	\label{Solution}
\end{figure}

Before presenting any performance results, the first indispensable step for any parallelization or optimization is to quantify the solution difference between the serial solution and the parallel solution. An iteratively converged double precision solution using 32 GPUs on HokieSpeed for a $256^3$ mesh is given in Fig.~\ref{Solution}. As can be seen, the solution is smooth everywhere and the relative error compared to the serial solution is just O($10^{-8}$), also seen in Section V of Ref~\cite{xue2018multi}. The small difference is due to round-off error accumulations on the GPUs. For all the optimizations presented in this paper, parallel solution correctness is always guaranteed.

\subsection{Scaling Performance Metrics}
Two basic metrics used in this paper are parallel speedup and efficiency. Speedup denotes how much faster the parallel version is compared to the serial version of the code, while efficiency represents how efficiently the processors are used. They are defined as follows, 
\begin{equation}
\label{speedup}
\mathrm{speedup}=\frac{t_{serial}}{t_{parallel}}
\end{equation}
\begin{equation}
\label{efficiency}
\mathrm{efficiency}=\frac{speedup}{np}
\end{equation}
where $np$ is the number of processors (CPUs or GPUs).

In order for the performance of the code to be measured and compared well on different platforms and for different problem sizes, the wall clock time per iteration step is converted to a ssspnt (scaled size steps per $np$ time) value which is defined in Eq.\ref{ssspnt}. This metric has some advantages. First, GFLOPS requires knowing the number of operations while ssspnt does not require. In most codes, it is usually difficult to know the number of operations. Second, efficiency comparison across different platforms is not intuitive as how fast the program runs is still unknown, but ssspnt is clearer for knowing the absolute speed. Also, the benefit applies to strong and weak scaling performance comparison if using ssspnt. For example, a linear scaling problem has a constant ssspnt value but the values can be different for the strong and weak scaling. Using ssspnt, different problems, platforms, strong and weak scaling performance can be compared directly. In this paper, when gaining productive performance results, unnecessary I/O (writing out solutions to files and writing residual norms to the screen) are turned off, which is commonly applied when testing the performance in literature.
\begin{equation}
\label{ssspnt}
\mathrm{ssspnt}=s\frac{size \times steps}{np \times time}
\end{equation}
where $s$ is a scaling factor which scales the smallest platform ssspnt to the range of [0,1]. In this paper, $s$ is set to be $10^{-7}$ for all test cases. $size$ is the problem size, $steps$ is the iteration steps and $time$ is the program wall clock time for $steps$ iterations.

\subsection{Grid Growth for Weak Scaling}
In the weak scaling analysis, the problem size needs to be increased accordingly when the number of processors increases. However, the way the problem scales can vary. For the BDC codes, since the problem is a 3D problem, the problem can increase in either 1D, or 2D, or 3D. Therefore, we will investigate the effect of how problem size grows on the weak scaling performance. Two types of grid growth for the weak scaling are applied, seen in Table~\ref{growth_type1} and Table~\ref{growth_type2}. Table~\ref{growth_type1} keeps the problem size grow in the exactly the same way when increasing the number of processors, no matter whether the codes uses 3D, 2D and 1D decomposition, and Table~\ref{growth_type2} scales the problem size in accordance with the way the number of processors grow. For example, if using 8 processors, the problem size for 3D decomposition is $512\times512\times512$, for 2D decomposition it is $256\times512\times1024$, and for 1D decomposition it is $256\times256\times2048$ or $2048\times256\times256$ (as the processor dims can be either $1\times1\times8$ or $8\times1\times1$). If not specified, the paper uses the grid growth in Table~\ref{growth_type2}.
\begin{table}[H]
	\caption{Grid growth type 1}
	\centering
	\label{growth_type1}
	\begin{tabular}{cccc}
		\hline
		Problem size & 3D decomposition & 2D decomposition & 1D decomposition                    \\ \hline
		(256,256,256)           & (1,1,1)               & (1,1,1)               & (1,1,1)               \\ 
		(256,256,512)           & (1,1,2)               & (1,1,2)               & (1,1,2) or (2,1,1)    \\ 
		(256,512,512)           & (1,2,2)               & (1,2,2)               & (1,1,4) or (4,1,1)    \\ 
		(512,512,512)           & (2,2,2)               & (1,2,4)               & (1,1,8) or (8,1,1)    \\ 
		(512,512,1024)          & (2,2,4)               & (1,4,4)              & (1,1,16) or (16,1,1)   \\ 
		(512,1024,1024)         & (2,4,4)               & (1,4,8)
		        & (1,1,32) or (32,1,1)   \\ \hline
	\end{tabular}
\end{table}

\begin{table}[H]
	\caption{Grid growth type 2}
	\centering
	\label{growth_type2}
	\begin{tabular}{cccc}
		\hline
		\# of processors & 3D decomposition & 2D decomposition & 1D decomposition                          \\ \hline
		1           & (256,256,256)               & (256,256,256)               & (256,256,256)               \\ \hline
		2           & (256,256,512)               & (256,256,512)               & \makecell{(256,256,512) or  \\ 
		                     (512,256,256)}    \\ \hline
		4           & (256,512,512)               & (256,512,512)               & \makecell{(256,256,1024) or \\
		                     (1024,256,256)}   \\ \hline
		8           & (512,512,512)               & (256,512,1024)              & \makecell{(256,256,2048) or \\
		                     (2048,256,256)}   \\ \hline
		16          & (512,512,1024)              & (256,1024,1024)             & \makecell{(256,256,4096) or \\
		                     (4096,256,256)}   \\ \hline
		32          & (512,1024,1024)             & (256,1024,2048)
		         & \makecell{(256,256,8192) or \\
		                     (8192,256,256)}   \\ \hline
	\end{tabular}
\end{table}

\subsection{Multi-CPU Scaling Performance}
\begin{figure}[H]
	\centering 
	\subfigure[CPU strong scaling]{ 
		\label{CPU_1D2D3Dstrong_ssspnt}
		\includegraphics[width=.45\textwidth]{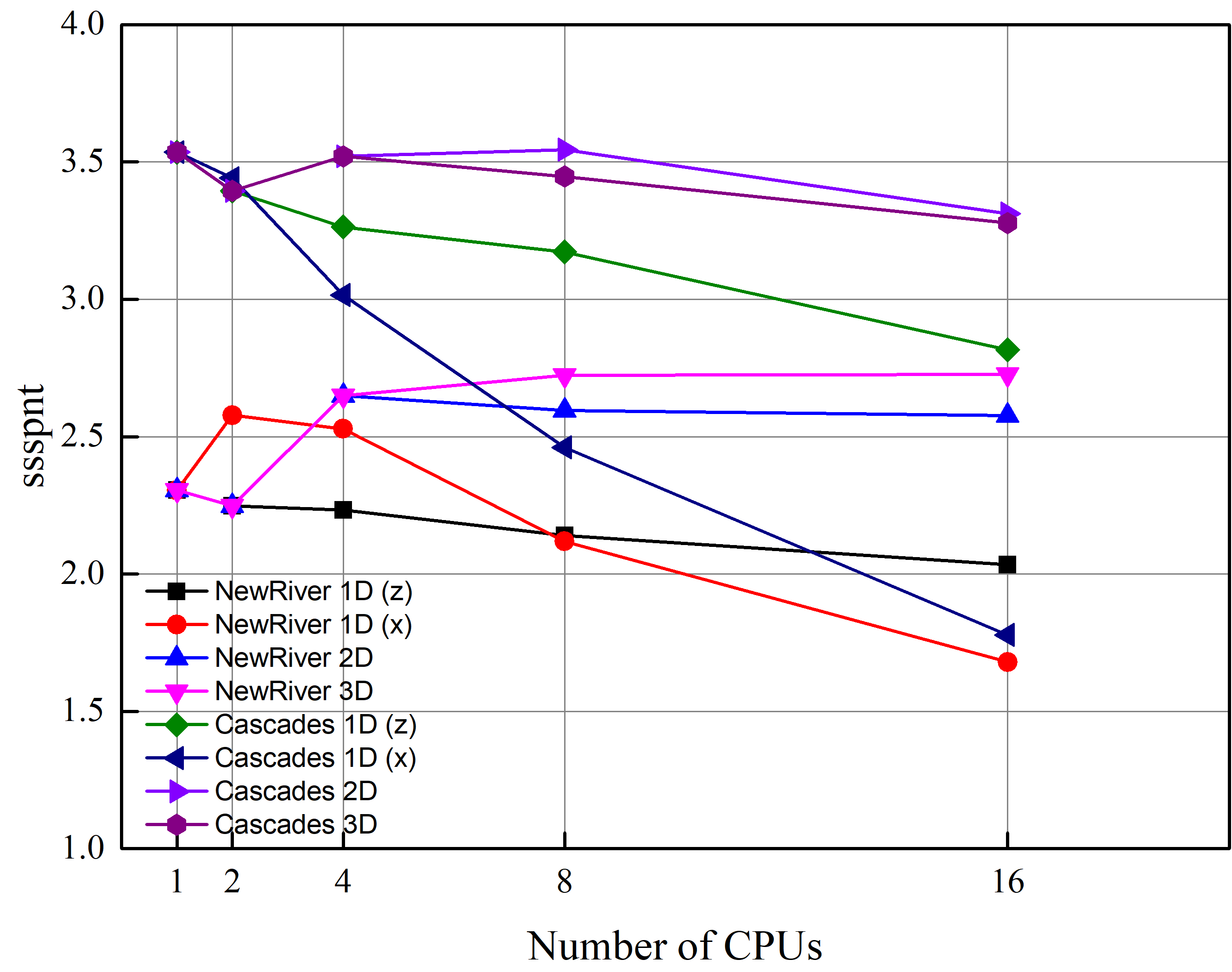} 
	} 
	\subfigure[CPU weak scaling]{ 
		\label{CPU_1D2D3Dweak_ssspnt}
		\includegraphics[width=.45\textwidth]{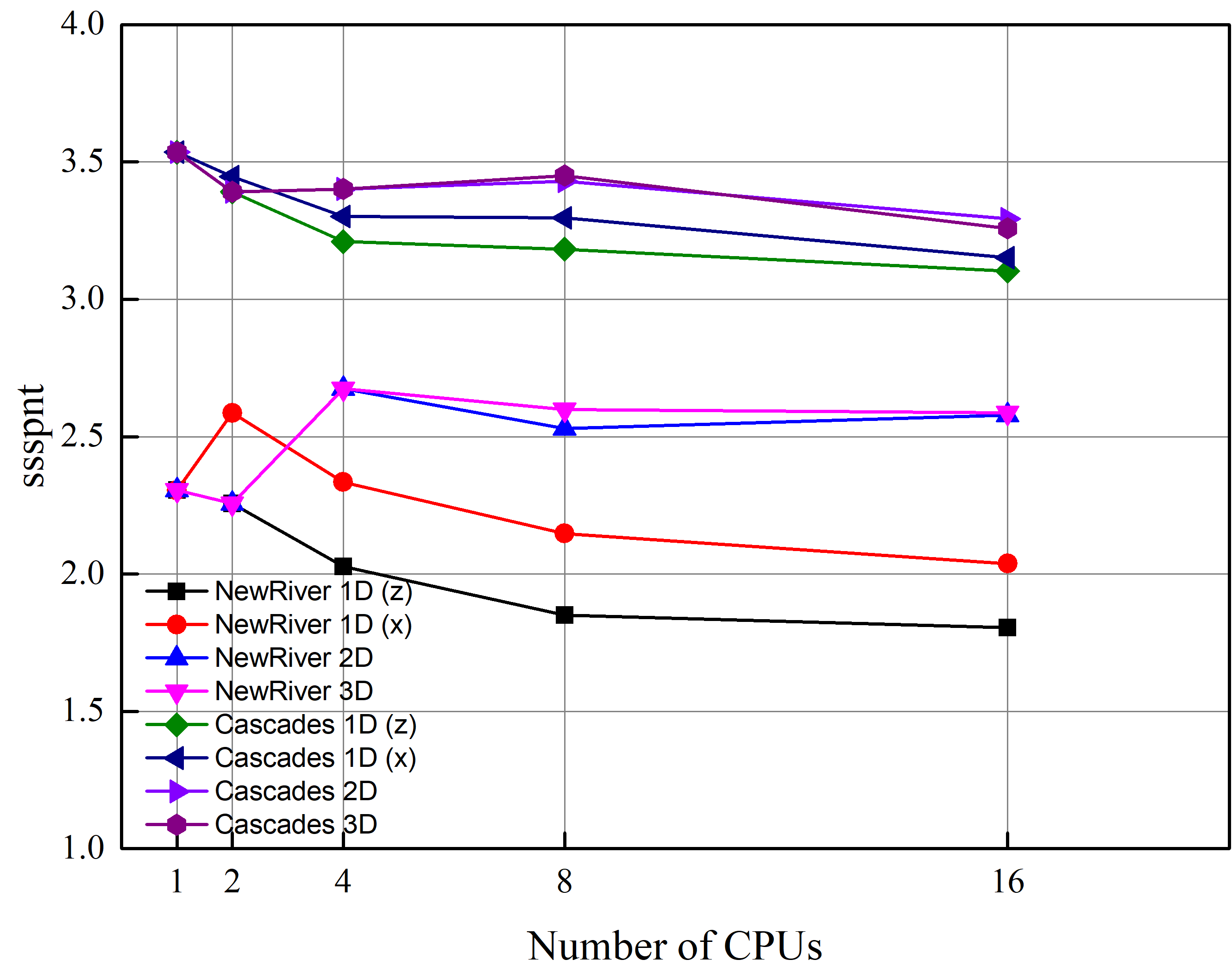} 
	} 
	\caption{Multi-CPU scaling using different decompositions} 
	\label{CPU_1D2D3D}
\end{figure}

A systematic multi-CPU scaling performance test is performed on all the three clusters mentioned earlier in this paper. CPU strong scaling and weak scaling performance using 1D, 2D and 3D decompositions are shown in Fig.~\ref{CPU_1D2D3D}. Here CPUs are added by sockets, i.e., adding a certain number of sockets every time (only using one CPU in each socket, similar to adding GPUs as every socket only has one GPU), or equivalently setting the processor per node (ppn) to be 2. Fig.~\ref{CPU_1D2D3D} highlights a tradeoff that exists for the different domain decomposition techniques. Choosing a domain decomposition scheme is a balance between maintaining a small surface to volume ratio of the subdomains and minimizing the number of neighbors for each subdomain. From Fig.\ref{CPU_1D2D3D}, 1D decomposition generally performs the worst for both the strong and weak scaling on NewRiver and Cascades. 1D decomposition has only 2 neighbors but it has the highest surface to volume ratio meaning that there is a lot of data to transfer between the blocks. 2D and 3D decomposition perform the best for the strong scaling and weak scaling, respectively. 3D decomposition has the smallest surface to volume ratio but has 6 neighbors meaning it has to perform 6 communications each with their own overhead. The weak scaling curves are flatter than the strong scaling, which is reasonable as the CPU has more work to do. It is also found that decomposing in the $x$ dimension should be avoided for the CPU in strong scaling as the performance deteriorates faster compared to other decompositions, which is especially obvious on the Cascades cluster.   

\begin{figure}[H]
	\centering 
	\subfigure[CPU strong scaling]{ 
		\label{CPU_strong_ssspnt_ppn}
		\includegraphics[width=.45\textwidth]{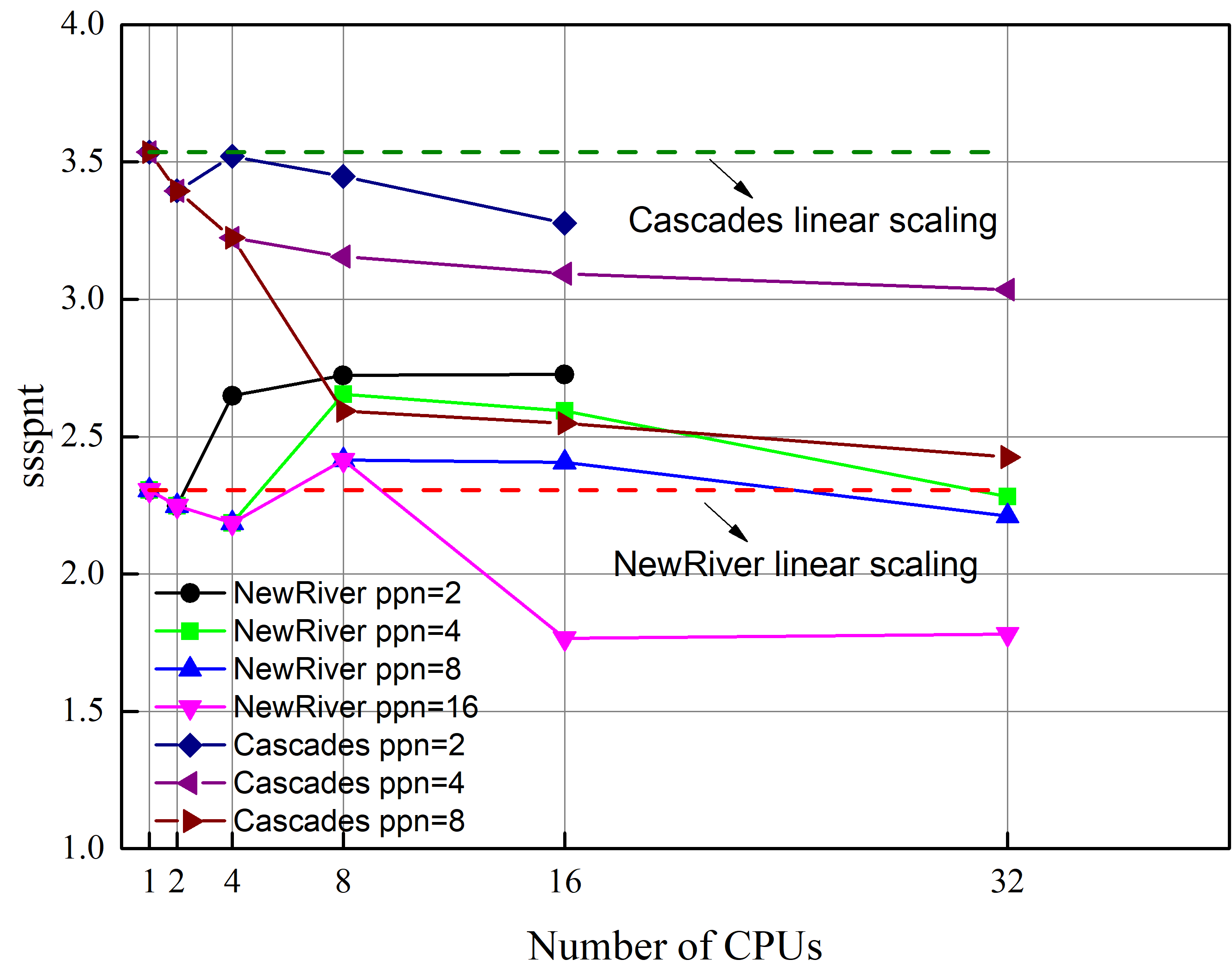} 
	} 
	\subfigure[CPU weak scaling]{ 
		\label{CPU_weak_ssspnt_ppn}
		\includegraphics[width=.45\textwidth]{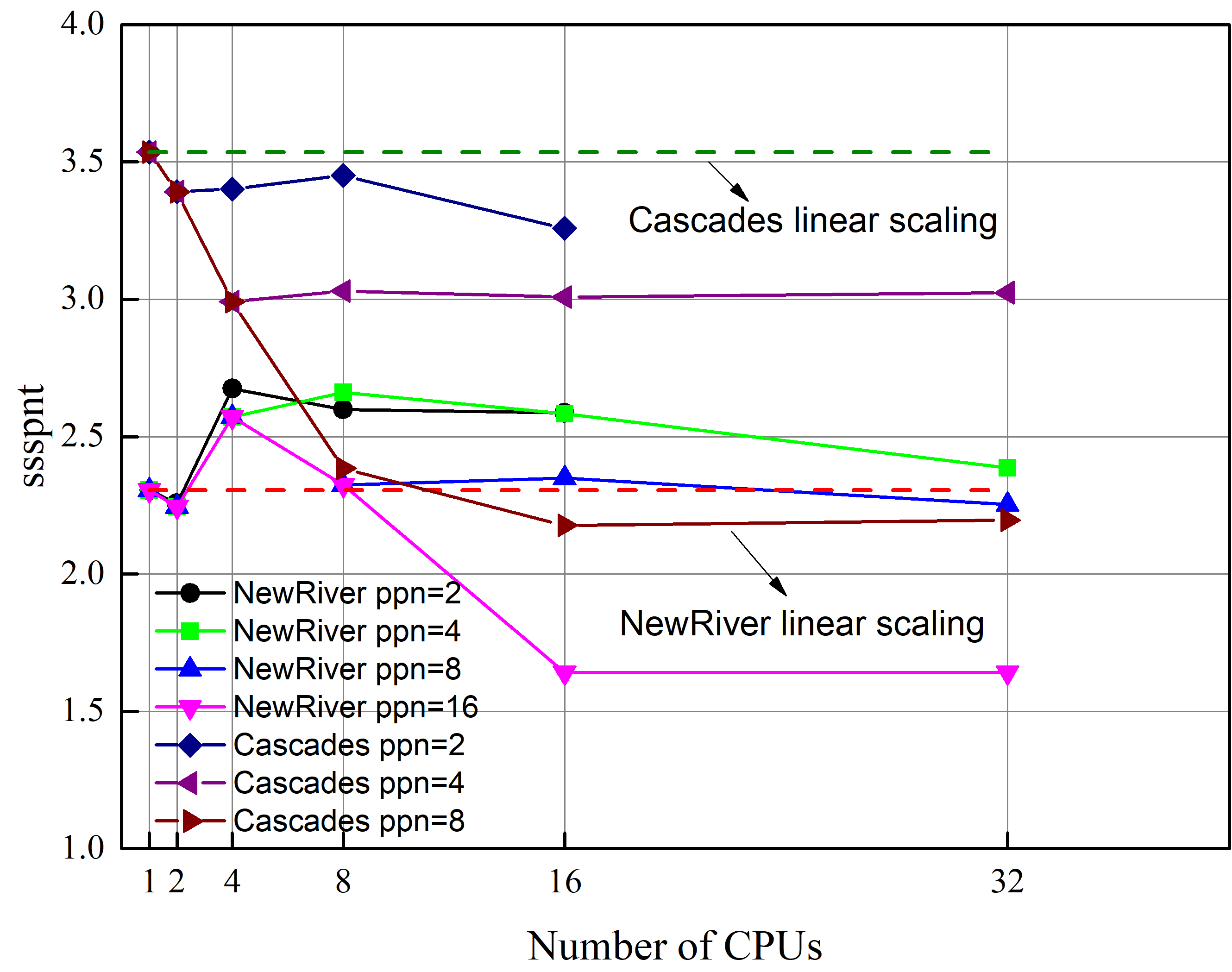} 
	} 
	\caption{The effect of ppn (3D decomposition)} 
	\label{CPU_ppn}
\end{figure}

Also, Fig.~\ref{CPU_1D2D3D} shows that a super-linear scaling occurs on the NewRiver cluster. To investigate why there is super-linear phenomenon, the ppn is changed. Fig.~\ref{CPU_ppn} shows the effect of ppn on the performance. For both the strong and weak scaling performance shown in Fig.~\ref{CPU_1D2D3D}, since the ppn is 2, the number of CPUs is increased along with other resources such as memory and memory bandwidth increased at the same pace (as the number of sockets increases), which may reduce the communication overhead and latency cost. To test this hypothesis, we also tried setting ppn to be higher than 2 (using less nodes) and did not observe the super-linear scaling. Also, no matter what the ppn is, the Cascades cluster does not show a super-linear scaling. It can be concluded that the super-linear phenomenon depends highly on the platform communication system and the implementation.

\begin{figure}[H]
	\centering
	\includegraphics[width=.48\textwidth]{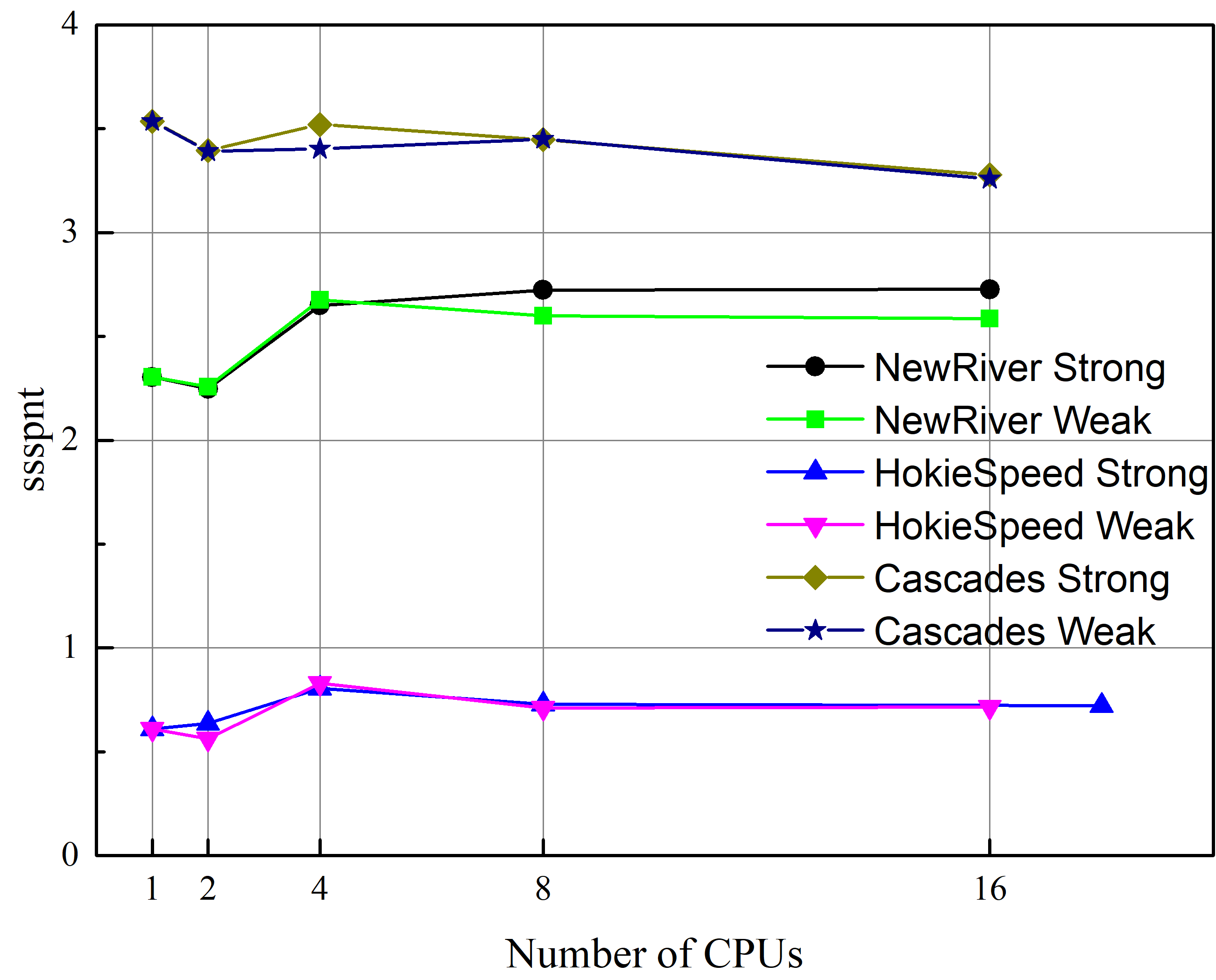}
	\caption{Multi-CPU performance comparison across platforms (3D domain decomposition)}
	\label{CPU_ssspnt_comparison}
\end{figure}

Although multi-CPU implementation is not the focus of this paper, we are interested in the CPU performance comparison between different platforms, which is shown in Fig.~\ref{CPU_ssspnt_comparison}. All the results here use 3D domain decomposition. It can be found that the platform affects the performance most, not the number of CPUs or whether the scaling is strong or weak. The performance on Cascades is about 1.245 times faster as that on NewRiver, which is very close to the clock rate ratio of 1.25 (3Ghz/2.4Ghz). For the CPU scaling, 3D domain decomposition maintains the efficiency very well so it is recommended.

\subsection{Multi-GPU Scaling Performance}\label{Multi-GPU}
The focus of this paper is the multi-GPU implementation and its performance optimizations. There are multiple optimizations of the GPU-accelerated CFD code in this subsection: the first version is a baseline code, which can be regarded as a naive GPU implementation based on the CPU code, and the other versions are incremental optimization versions, with more optimizations based on the previous version. Actual memory bandwidth can be improved greatly as well communication overhead can be reduced by applying these optimizations. When showing results later, the "1D", "2D" and "3D" in the legend denotes 1D, 2D and 3D decompositions, respectively. For 1D decomposition, the letter in the parentheses after "1D" denotes which dimension to decompose. More details about these different GPU-accelerated versions are given as follows. 

\paragraph{Baseline}
The multi-CPU code is directly ported to GPUs by inserting OpenACC directives in the parallel CPU code. This baseline GPU code does not use GPUDirect techniques. Therefore, data on devices need to be updated to/from hosts using "\texttt{!\$acc update}" clauses. Asynchronization clauses are used to reduce some synchronization overhead between hosts and devices. 

\paragraph{Optimized V1}
This version is to improve the actual memory bandwidth and reduce latency cost using 3D decomposition, as we found that the actual bandwidth between the host and the device is very small because of non-contiguous data transfer. As Fortran is a column-majored language, the first index $i$ of a matrix $A(i,j,k)$ denotes the fastest change. If any decomposition exists in the $i$ index direction (3D decomposition or 1D decomposition in $i$), the decomposition in the $i$ index direction can generate chunks of data (at $j-k$ planes) which are highly non-contiguous. Therefore, the optimization is targeted at solving this issue by converting the non-contiguous data into a temporary contiguous array in parallel and updating this temporary array between hosts and devices using OpenACC update clauses. For this optimization, temporary arrays are created only if decomposition in the direction with the fastest change ($i$ index direction) exists, as decomposition in other index directions can still generate chunks of contiguous data. A pseudo code of how to use buffers is given in Listing.~\ref{Pseudo}. The procedure can be summarized as follows:
\begin{enumerate}
\item Allocate send/recv buffers only for boundary cells on $i$ planes on devices and hosts if decomposition happens in the $i$ dimension, as the non-contiguous data on $i$ planes makes data transfer very slow.
\item Pack the noncontiguous block boundary data to a buffer on the sender device side. This process can be parallelized using "\texttt{!\$acc loop}" clauses and has little overhead. The host buffer is then updated using OpenACC update host clauses.
\item Have hosts transfer the data through MPI\_Isend/MPI\_Irecv calls (which are one-sided non-blocking calls). Then block MPI calls using MPI\_WAITALL to finish data transfer.
\item Update the recv buffer on devices using OpenACC update device clauses and finally unpack the contiguous data stored in recv buffer back to noncontiguous memory on devices, which can also be parallelized using "\texttt{!\$acc loop}" clauses.
\end{enumerate}

\begin{lstlisting}[caption={A pseudo code of optimization on non-contiguous data transfer between hosts and devices},label=Pseudo, language=Fortran]
!$acc data present(send_buffer(:,:), soln(:,:,:,:))

!Pack send_buffer(:,1) with back boundary data
!$acc parallel loop collapse(4) async(1)
do var = 1, 5
  do k= 1, k_nodes
    do j= 1, j_nodes
      !Pack up two layers of cells
      do i = 1, 2
        indx = var*k_nodes*j_nodes*2+k*j_nodes*2+j*2+i
        !Interior cell index starts at 3
        send_buffer(indx,1) = soln(i+2,j,k,var)
        
!Similar routine to pack send_buffer(:,2) with front boundary data async(2)

!Update send_buffer on hosts
!$acc update host(send_buffer)
!$acc wait

!Send/Recv between hosts
MPI_IRECV(recv_buffer)
MPI_ISEND(send_buffer)
MPI_WAITALL

!Update recv_buffer on devices
!$acc update device(recv_buffer)
!$acc wait

!$acc data present(recv_buffer(:,:), soln(:,:,:,:))

!Unpack the data in recv_buffer to soln ghost locations
\end{lstlisting}

\paragraph{Optimized V2}
The use of only one stencil in the \emph{Optimized V1} makes the communication pattern and implementation simpler but may not be efficient. \emph{Optimized V2} is designed to reduce the amount of data exchanged. Since the pressure requires a larger stencil while other primitive variables do not, We can transfer less less data based on their own stencil size compared to \emph{Optimized V1}. What is more, more overlap of asynchronous communication can be achieved as a big loop is split into two asynchronous loops. A pseudo code of this optimization is given in Listing.~\ref{Pseudo_Stencil}. It should be noted that only changes based on the previous optimization are emphasized in a new Listing. 
\begin{lstlisting}[caption={A pseudo code of stencil based communication optimization},label=Pseudo_Stencil, language=Fortran]
!$acc data present(send_buffer(:,:), soln(:,:,:,:))

!Pack send_buffer(:,1) with back boundary data
!Move pressure into send_buffer(:,1)
!$acc parallel loop collapse(3) async(1)
do k= 1, k_nodes
  do j= 1, j_nodes
    do i = 1, 2
      var = 1 ! pressure only
      indx = k*j_nodes*2+j*2+i
      send_buffer(indx,1) = soln(i+2,j,k,var)
      
!Update starting index in send_buffer(:,1)
indx_p = k_nodes*j_nodes*2

!Move velocities & temperature into send_buffer(:,1)       
!$acc parallel loop collapse(3) async(2)
do var = 2, 5
  do k= 1, k_nodes
    do j= 1, j_nodes
      indx = (var-2)*k_nodes*j_nodes+k*j_nodes+j
      send_buffer(indx_p+indx) = soln(3,j,k,var)

!Pack send_buffer(:,2) with front boundary data
!$acc parallel loop collapse(3) async(3) & async(4)

!Update send_buffer on hosts
...

!Send/Recv between hosts
...

!Update recv_buffer on devices
...

!$acc data present(recv_buffer(:,:), soln(:,:,:,:))

!Unpack the data in recv_buffer to soln ghost locations async(1:4)
\end{lstlisting}

\paragraph{Optimized V3}
In the \emph{Optimized V1} and \emph{Optimized V2}, contiguous-memory arrays are created only for the $i$ index direction. However, if an decomposition exists in the $j$ or $k$ index direction, then we may also need an array in the $j$ and $k$ direction. It should be noted that although real cell data on $k$ boundary faces do not need to be packed into buffers, using buffers on $k$ faces may still be helpful considering that there are ghost cells on $k$ boundary faces which breaks the contiguity. We found that on the HokieSpeed cluster, using such arrays improves the performance very little, but the performance can be improved significantly on the NewRiver and Cascades cluster. The procedure of creating arrays and parallelizing the pack/unpack process in the $j$ and $k$ direction is very similar to that in the $i$ direction, so there is no need to show an pseudo code here. Readers can reference Listing.~\ref{Pseudo}.

\begin{figure}[H]
	\centering
	\includegraphics[width=.48\textwidth]{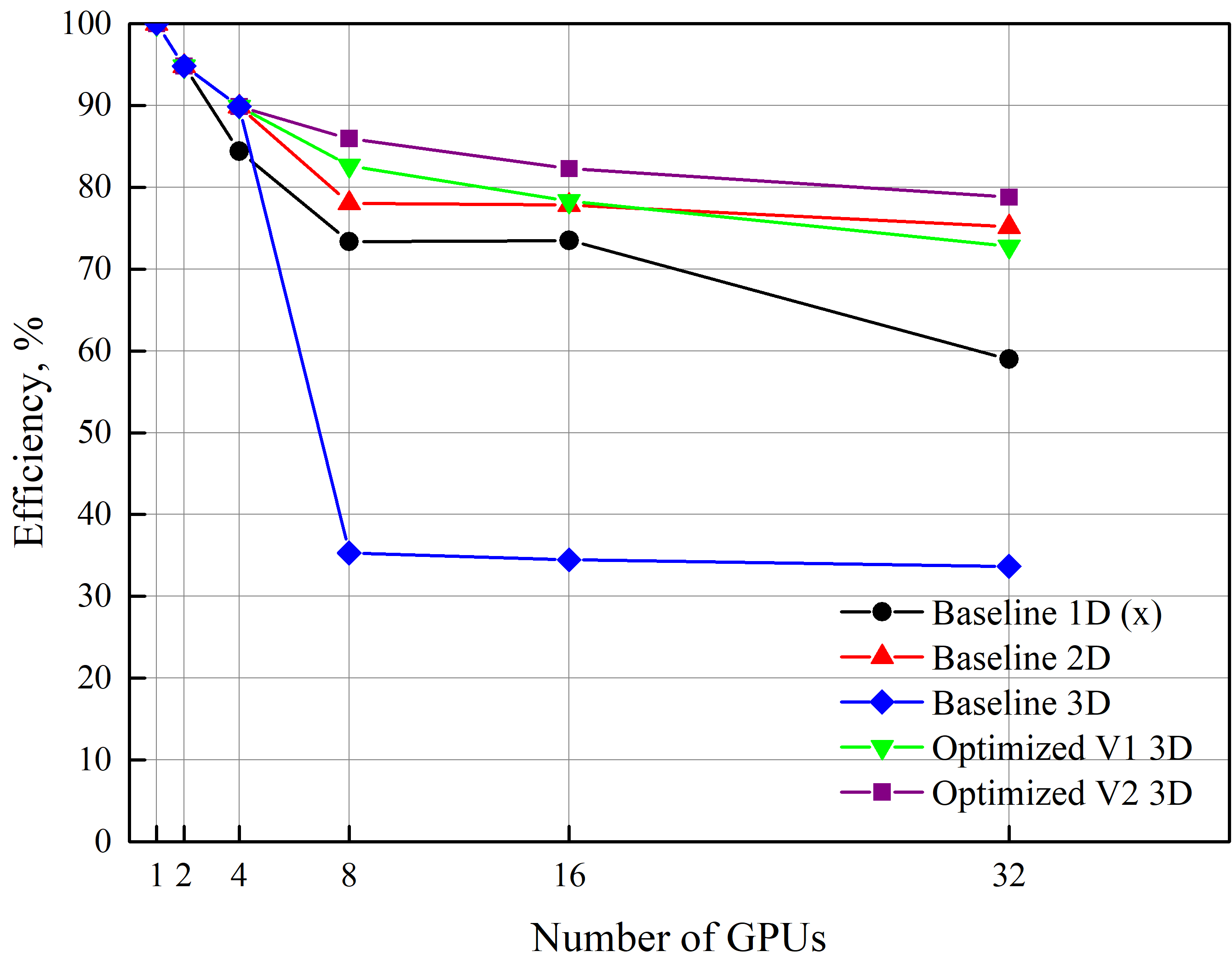}
	\caption{Multi-GPU scaling performance on the HokieSpeed cluster}
	\label{GPU_hokiespeed_weak}
\end{figure}

We will first show the benefit of applying \emph{Optimized V1} on the HokieSpeed cluster, i.e., creating the contiguous-memory and parallelizing the process of pack/unpack. Fig.~\ref{GPU_hokiespeed_weak} shows the weak scaling efficiency of the different GPU code versions introduced earlier. The \emph{Baseline} version using 3D decomposition performs poorly, which is very bandwidth limited due to noncontiguous data transfer when 3D decomposition is used. Both \emph{Baseline} 2D and 1D perform much better than \emph{Baseline} 3D. Although the baseline GPU version using 3D decomposition performs poorly, its two optimizations scales as well as the baseline using 2D decomposition or even better. This indicates that memory throughput is improved greatly and latency cost is reduced after optimization, although there is some pack/unpack overhead. Therefore, special attention should be paid to non-contiguous data movement between hosts and devices. Also, Fig.~\ref{GPU_hokiespeed_weak} shows that \emph{Optimized V2} performs better than \emph{Optimized V1} because it transfers less data and overlaps asynchronous data transfers better. Similar performance improvement can also be seen on the NewRiver and Cascades clusters but the baseline scaling performance are not shown in this paper, as the baseline version runs too slow (or it can be regarded as an improper implementation) and its performance assessment is not a focus in this paper.

\begin{figure}[H]
	\centering 
	\subfigure[GPU strong scaling]{ 
		\label{GPU_1D2D3Dstrong_ssspnt}
		\includegraphics[width=.45\textwidth]{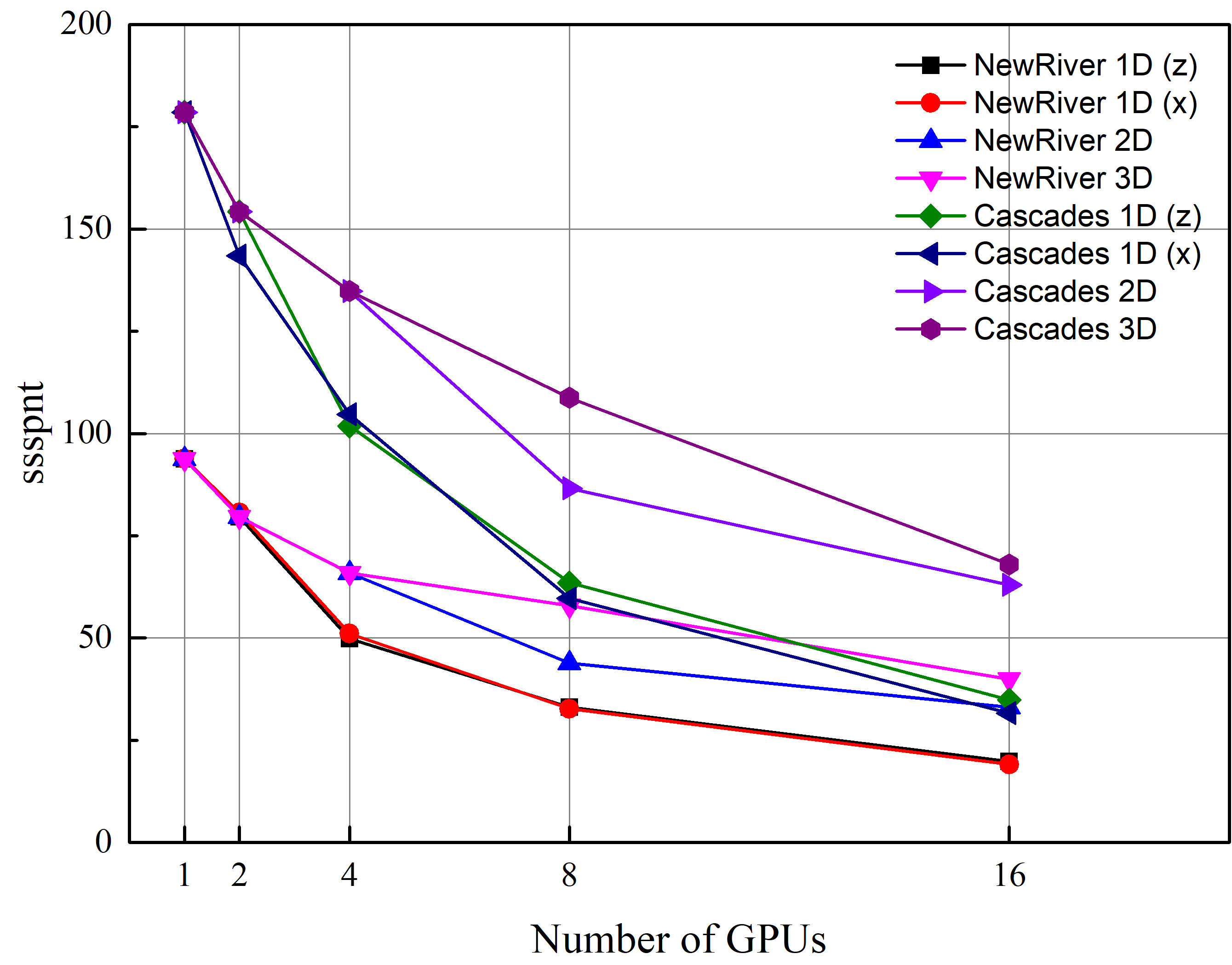} 
	} 
	\subfigure[GPU weak scaling]{ 
		\label{GPU_1D2D3Dweak_ssspnt}
		\includegraphics[width=.45\textwidth]{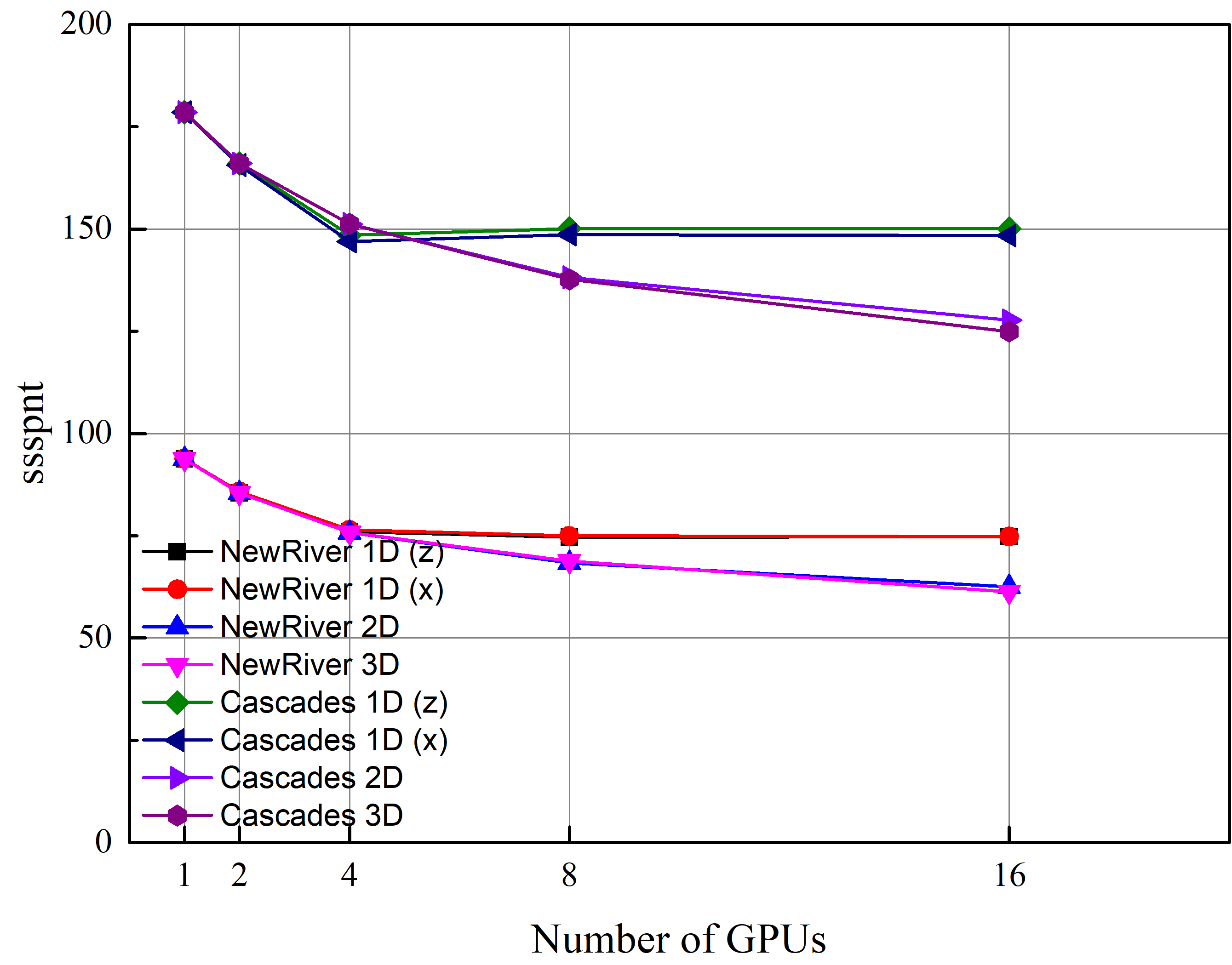} 
	} 
	\caption{Multi-GPU scaling using different decompositions (\emph{Optimized V3})} 
	\label{GPU_1D2D3D}
\end{figure}

From Fig.~\ref{GPU_hokiespeed_weak}, we know that \emph{Optimized V2} performs the best on the HokieSpeed cluster. It should be mentioned that \emph{Optimized V2} and \emph{Optimized V3} perform equivalently on HokieSpeed. However, NewRiver and Cascades show different favors. The performance comparison of \emph{Optimized V2} and \emph{Optimized V3} on NewRiver and Cascades will be shown in Sec~\ref{Sec_GPUDirect}, as there the performance of applying different MPI compilers and GPUDirect will include such an comparison. To avoid redundant contents, only the \emph{Optimized V3} performance on NewRiver and Cascades are shown here. The GPU strong scaling and weak scaling performance using 1D, 2D and 3D decompositions are shown in Fig.~\ref{GPU_1D2D3D}. From Fig.~\ref{GPU_1D2D3D}, 3D decomposition performs the best for the strong scaling, then 2D decomposition follows. 1D decomposition in the $x$ or $z$ dimension makes the performance drop quickly, especially on Cascades. Similar to the CPU weak scaling, the weak scaling curves are flatter than strong scaling. Recall that this weak scaling applies the grid growth type 2, which is in Table~\ref{growth_type2}. Since the problem size increases in accordance with the way $np$ grows, 1D decomposition performs the best as every decomposed block has the least number of neighbours compared to 2D and 3D decomposition.

\begin{figure}[H]
	\centering
	\includegraphics[width=.48\textwidth]{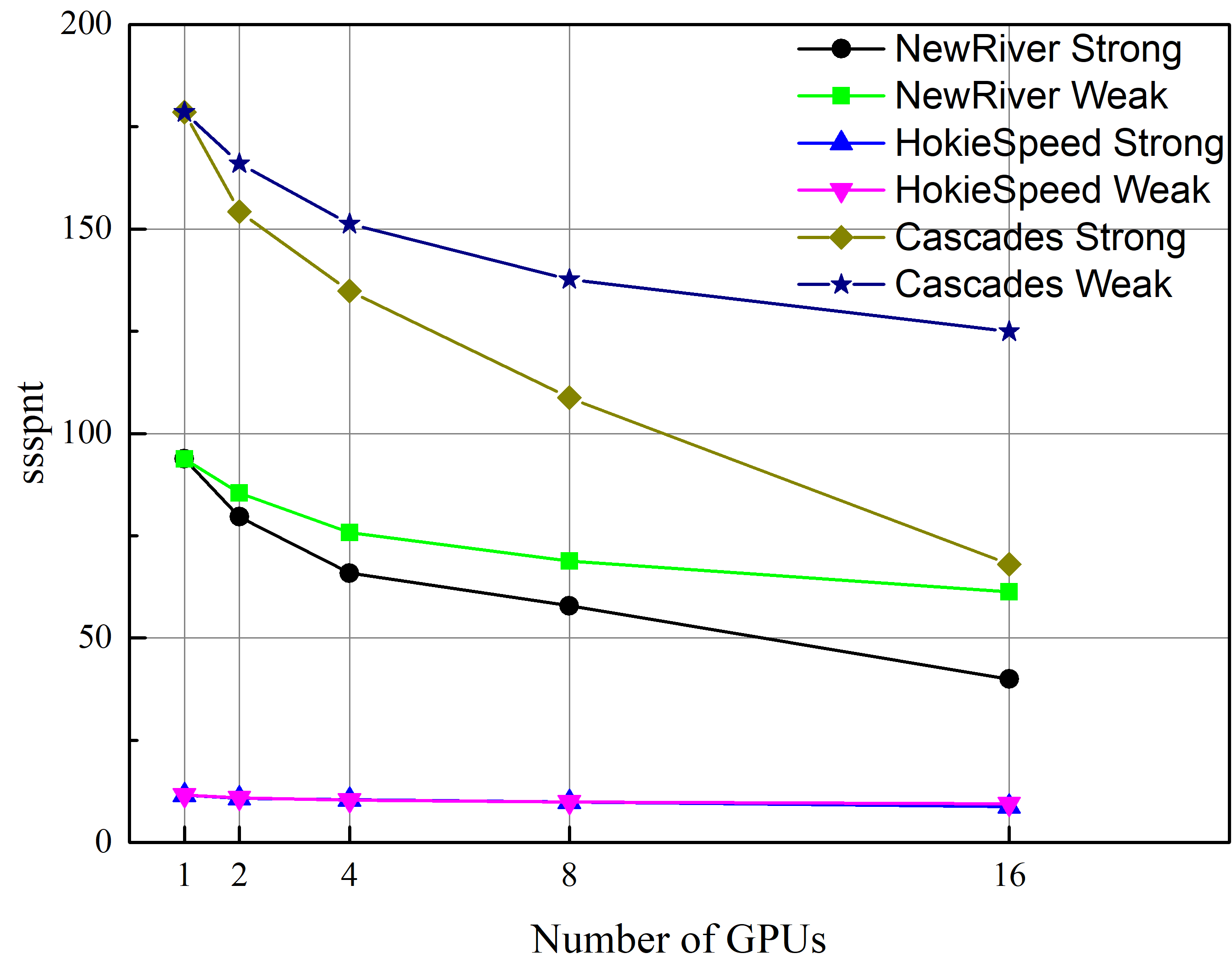}
	\caption{Multi-GPU performance comparison across platforms (\emph{Optimized V3})}
	\label{GPU_ssspnt_comparison}
\end{figure}

Fig.~\ref{GPU_ssspnt_comparison} shows the performance comparison across platforms. Different from Fig.~\ref{CPU_ssspnt_comparison}, multiple factors can affect the multi-GPU performance significantly, including the number of processors, platforms, whether a strong or weak scaling. When the number of GPUs increases, the efficiency drops significantly for both the strong and weak scaling, but the weak scaling efficiency holds a relatively higher value compared to the strong scaling. Cascades shows an about 2 times faster speedup compared to NewRiver.

\begin{figure}[H]
	\centering
	\includegraphics[width=.48\textwidth]{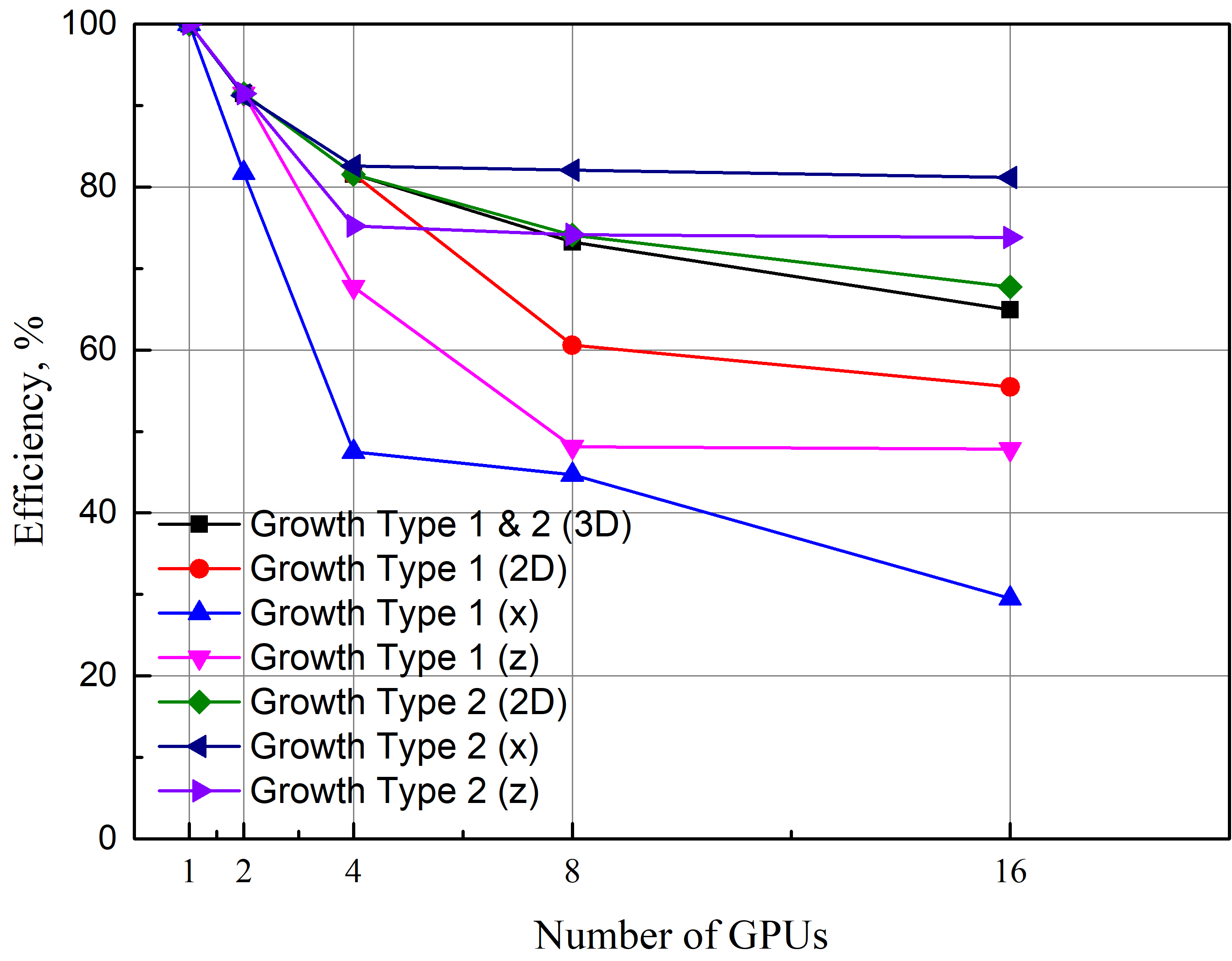}
	\caption{Weak scaling performance applying different grid growth methods (\emph{Optimized V3})}
	\label{GPU_grid_growth_comparison}
\end{figure}

To investigate the effect of different grid growth methods on the weak scaling performance, some cases are tested on NewRiver and Fig.~\ref{GPU_grid_growth_comparison} shows such results. Since the two grid growths are the same if applying 3D decomposition, there is only one curve for 3D decomposition. It can be found that the performance is better for growth type 2. If growth type 1 is applied, then 3D decomposition performs the best, and it may mislead readers that 3D decomposition is the best for the GPU weak scaling, which is not this paper's intention. It should be emphasized again that which decomposition should be used for the weak scaling depends on how the grid grows.

\subsection{CUDA-aware MPI and GPUDirect}\label{Sec_GPUDirect}
The optimizations introduced in Section~\ref{Multi-GPU} have improved the efficiency significantly on different platforms. However, we are still interested in improving the scaling performance further on NewRiver and Cascades since they have some modern GPUs. Thus, it became important to determine ways of reducing this communication cost by using CUDA-aware MPI and GPUDirect~\cite{CUDA_MPI}. Later, performance comparisons will be made between using Open MPI, MVAPICH2 or MVAPICH2-GDR with GPUDirect.

HokieSpeed does not support CUDA-aware MPI. Thus, all inter-node GPU communications on HokieSpeed had to go through host memory. This staging deteriorates the performance greatly. Using CUDA-aware MPI, we only need to send GPU buffers instead of CPU buffers. CUDA-aware MPI has two performance benefits~\cite{CUDA_MPI}. First, operations which require message transfer can be pipelined, which improves the memory throughput. Second, acceleration techniques such as GPUDirect can be utilized by the MPI library transparently to the user.

GPUDirect is an umbrella word for several GPU communication acceleration technologies. It provides high bandwidth and low latency communication between NVIDIA GPUs. There are three levels of GPUDirect~\cite{GPUDirect}. The first level is GPUDirect Shared Access, introduced with CUDA 3.1. This feature avoids an unnecessary memory copy within host memory between the intermediate pinned buffers of the CUDA driver and the network fabric buffer. The second level is GPUDirect Peer-to-Peer transfer (P2P transfer) and Peer-to-Peer memory access (P2P memory access), introduced with CUDA 4.0. This P2P memory access allows buffers to be copied directly between two GPUs on the same node. The last is GPU RDMA (Remote Direct Memory Access), with which buffers can be sent from the GPU memory to a network adapter without staging through host memory. The last feature is not supported on NewRiver and Cascades as it pertains to specific versions of the drivers (both from NVIDIA for the GPU and Mellanox for the Infiniband) which are not installed (other dependencies exist on NewRiver and Cascades, particularly parallel filesystems). Although GPU RDMA is not available, the other aspects of GPUDirect can be tested to determine its effect on the scaling performance using MVAPICH2-GDR.

\subsubsection{Intra-Node Scaling Performance Results}  
In this subsection, we will first show the benefits of applying GPUDirect in a node. Table~\ref{GPUDirect_Strong} shows the strong scaling performance comparison of different GPU code versions using 2 GPUs (intra-node performance) on NewRiver. The problem size is $256^3$. The versions defined here are similar to the versions introduced in Section.~\ref{Multi-GPU}, with the \emph{Baseline} to be the non-optimized GPU version, \emph{Optimized V3} uses the pack/unpack in all the available dimensions and the stencil-based communication method, and GPUDirect uses the P2P transfer technology applied to both of these versions of the code. Within a node, we are using GPUDirect P2P transfer between the memory of two GPUs on the same system/PCIe bus.
\begin{table}[H]
	\caption{Strong scaling comparison of different GPU code versions using 2 GPUs (NewRiver)}
    \centering
    \label{GPUDirect_Strong}
	\begin{tabular}{cccc}
		\hline
		GPU code versions        & Decompositions & ssspnt & 
		Efficiency                              \\ \hline
		Single GPU               & (1,1,1)        &  93.8              & 100\%                              \\
		Baseline                 & (1,1,2)        & 145.9              & 77.8\%                             \\
		Baseline                 & (2,1,1)        &  22.0              & 11.7\%                             \\
		Optimized V3             & (1,1,2)        & 167.2              & 89.2\%                             \\
		Optimized V3             & (2,1,1)        & 169.8              & 90.5\%                             \\
		Baseline + GPUDirect     & (1,1,2)        & 155.8              & 83.1\%                             \\
		Baseline + GPUDirect     & (2,1,1)        & 154.7              & 82.5\%  		                    \\
		Optimized V3 + GPUDirect & (1,1,2)        & 177.9              & 94.9\%                             \\
		Optimized V3 + GPUDirect & (2,1,1)        & 179.4              & 95.7\%                             \\ \hline
	\end{tabular}
\end{table}

Using MVAPICH2, the baseline code decomposed in the $i$ direction performs poorly, about $1/7$ of that decomposed in the $k$ direction. After a series of optimizations the ssspnt value changes from 22.0 to 169.8, indicating again the importance of the coalesced memory access when doing host-device transfers. GPU direct P2P transfer on the baseline code is also able to avoid the cost of host-device transfers and is able to maintain an efficiency of $83\%$ even though the data is not contiguous. Combining the performance optimizations with the use of GPUDirect can improve the efficiency to approximately $95\%$ on 2 GPUs.

Table.~\ref{GPUDirect_Weak} shows the weak scaling performance comparison of different GPU code versions using 2 GPUs in the intra-node mode. The result also shows either the optimizations proposed in this paper or GPUDirect (or both if applicable) should be used, if non-contiguous data transfers happen. It is also reasonable to see that the weak scaling generally performs better than the strong scaling.
\begin{table}[H]
	\caption{Weak scaling comparison of different GPU code versions using 2 GPUs (NewRiver)}
	\centering
	\label{GPUDirect_Weak}
	\begin{tabular}{cccc}
		\hline
		GPU code versions        & Decompositions & ssspnt & 
		Efficiency                              \\ \hline
		Single GPU               & (1,1,1)        &  93.8              & 100\%                              \\
		Baseline                 & (1,1,2)        & 161.6              & 86.1\%                             \\
		Baseline                 & (2,1,1)        &  21.4              & 11.4\%                             \\
		Optimized V3             & (1,1,2)        & 175.3              & 93.5\%                             \\
		Optimized V3             & (2,1,1)        & 176.3              & 94.0\%                             \\
		Baseline + GPUDirect     & (1,1,2)        & 167.9              & 89.5\%                             \\
		Baseline + GPUDirect     & (2,1,1)        & 154.9              & 82.6\%  		                    \\
		Optimized V3 + GPUDirect & (1,1,2)        & 180.0              & 96.0\%                             \\
		Optimized V3 + GPUDirect & (2,1,1)        & 180.9              & 96.5\%                             \\ \hline
	\end{tabular}
\end{table}

\subsubsection{Inter-Node Scaling Performance Results}

\paragraph{Strong Scaling Performance Results}

\begin{figure}[H]
	\centering 
	\subfigure[Open MPI]{ 
		\label{GPU_openmpi_strong}
		\includegraphics[width=.30\textwidth]{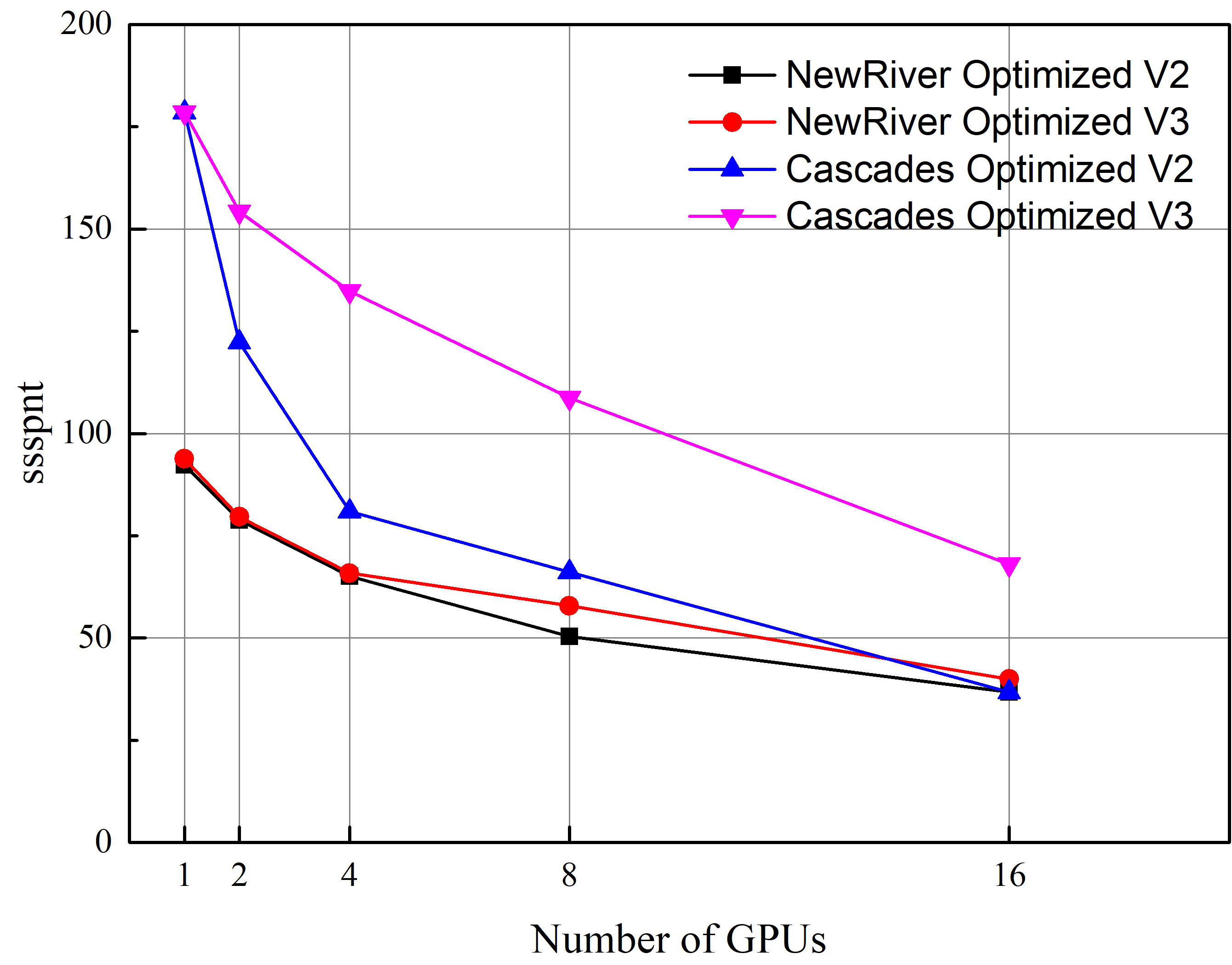} 
	}
	\subfigure[MVAPICH2]{ 
		\label{GPU_mvapich2_strong}
		\includegraphics[width=.30\textwidth]{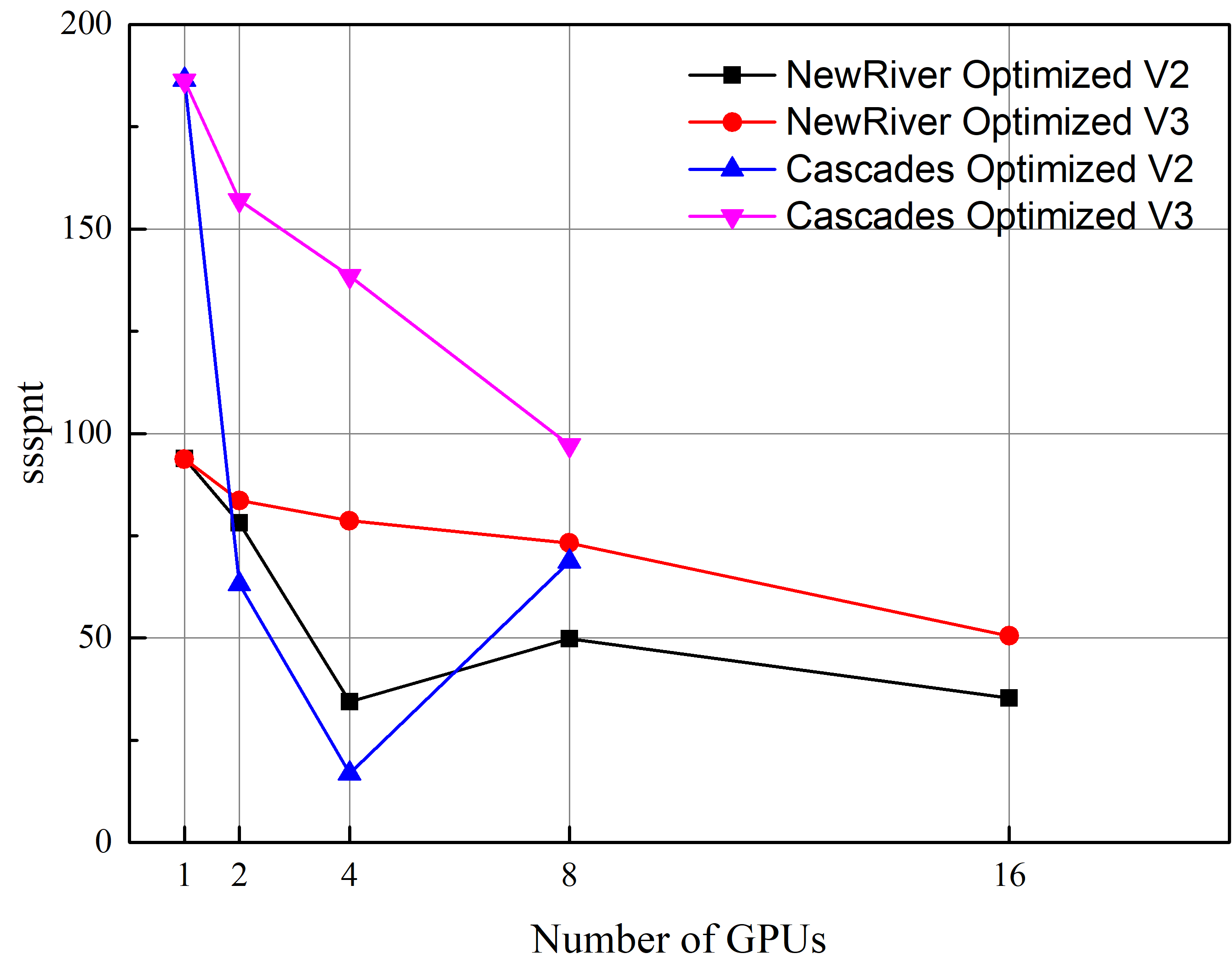}
	}
	\subfigure[MVAPICH2-GDR]{ 
		\label{GPU_mvapich2_gdr_strong}
		\includegraphics[width=.30\textwidth]{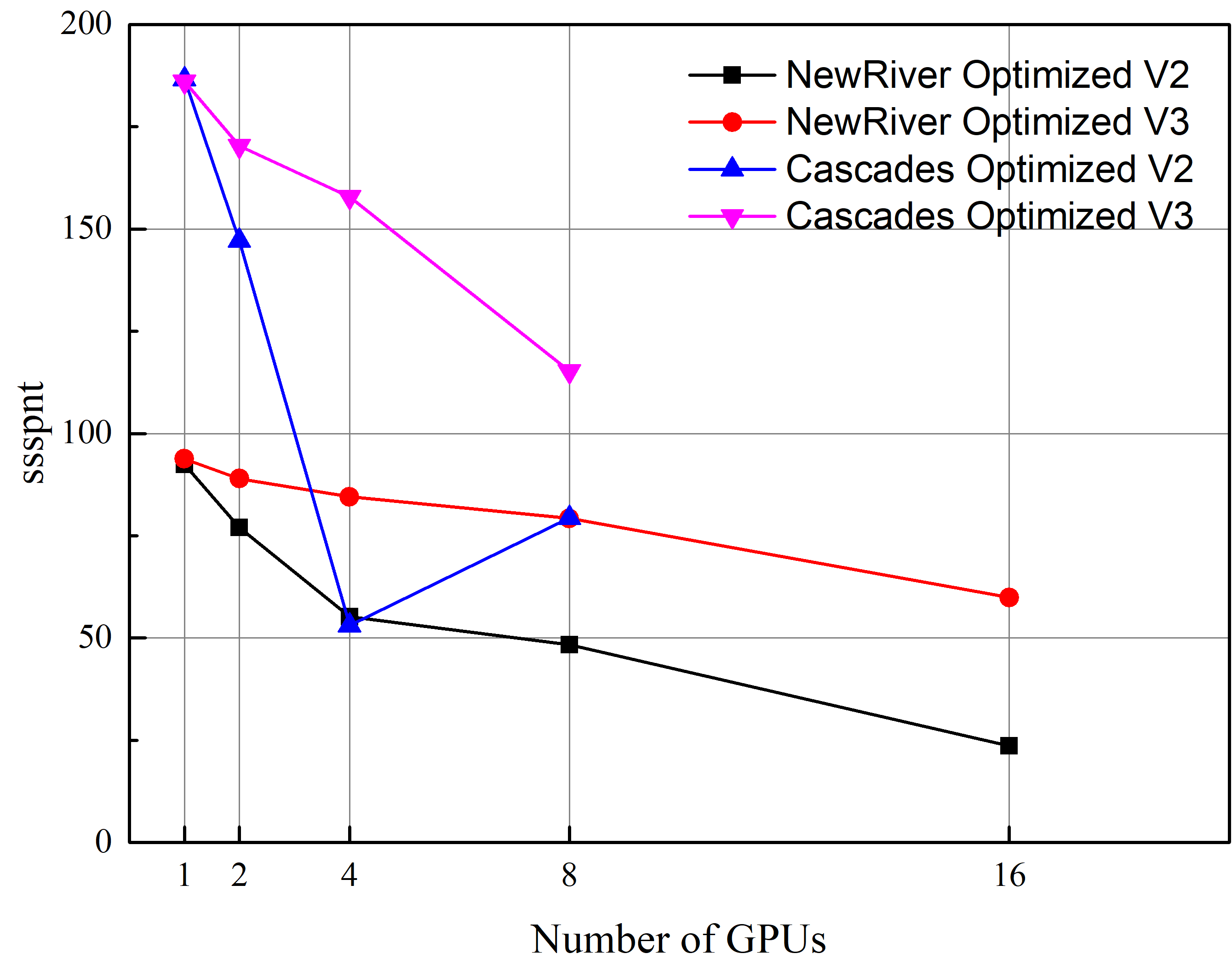}
	}	
	\caption{Strong scaling performance across platforms (3D decomposition)} 
	\label{GPUDirect_strong}
\end{figure}

Since there are three different MPI options (Open MPI, MVAPICH2 and MVAPICH2-GDR with GPUDirect turned on) on NewRiver and Cascades, scaling performance results using the three different compilers/options are given. Fig.~\ref{GPUDirect_strong} shows the strong scaling performance using different MPI options, respectively. Considering 3D growth is much more common in CFD such as applying systematic mesh refinement so the 3D decomposition is of more interest. As mentioned earlier, when applying MVAPICH2 or MVAPICH2-GDR on Cascades, the performance drops to 1\% if using 16 GPUs, so the maximum number of GPUs used in that occasion is 8. Since the results are for the strong scaling, we cannot expect a very high efficiency if scaling up to a large number of GPUs. Using MVAPICH2-GDR generally achieves the best performance especially when combined with \emph{Optimized V3} but it shows a significant performance drop when using 4 GPUs, which is more obvious on Cascades. The performance curves using Open MPI on both platforms are much smoother than using MVAPICH2 and MVAPICH2-GDR.

Also, a comparison can be made between \emph{Optimized V2} and \emph{Optimized V3} to address the importance of using more buffers for multi-GPU computing on modern GPUs. Since \emph{Optimized V3} generally uses more transfer buffers than \emph{Optimized V2}, the performance can be much better if decomposition exists in more dimensions.

\paragraph{Weak Scaling Performance Results}

\begin{figure}[H]
	\centering 
	\subfigure[OpenMPI]{ 
		\label{GPU_openmpi_weak}
		\includegraphics[width=.30\textwidth]{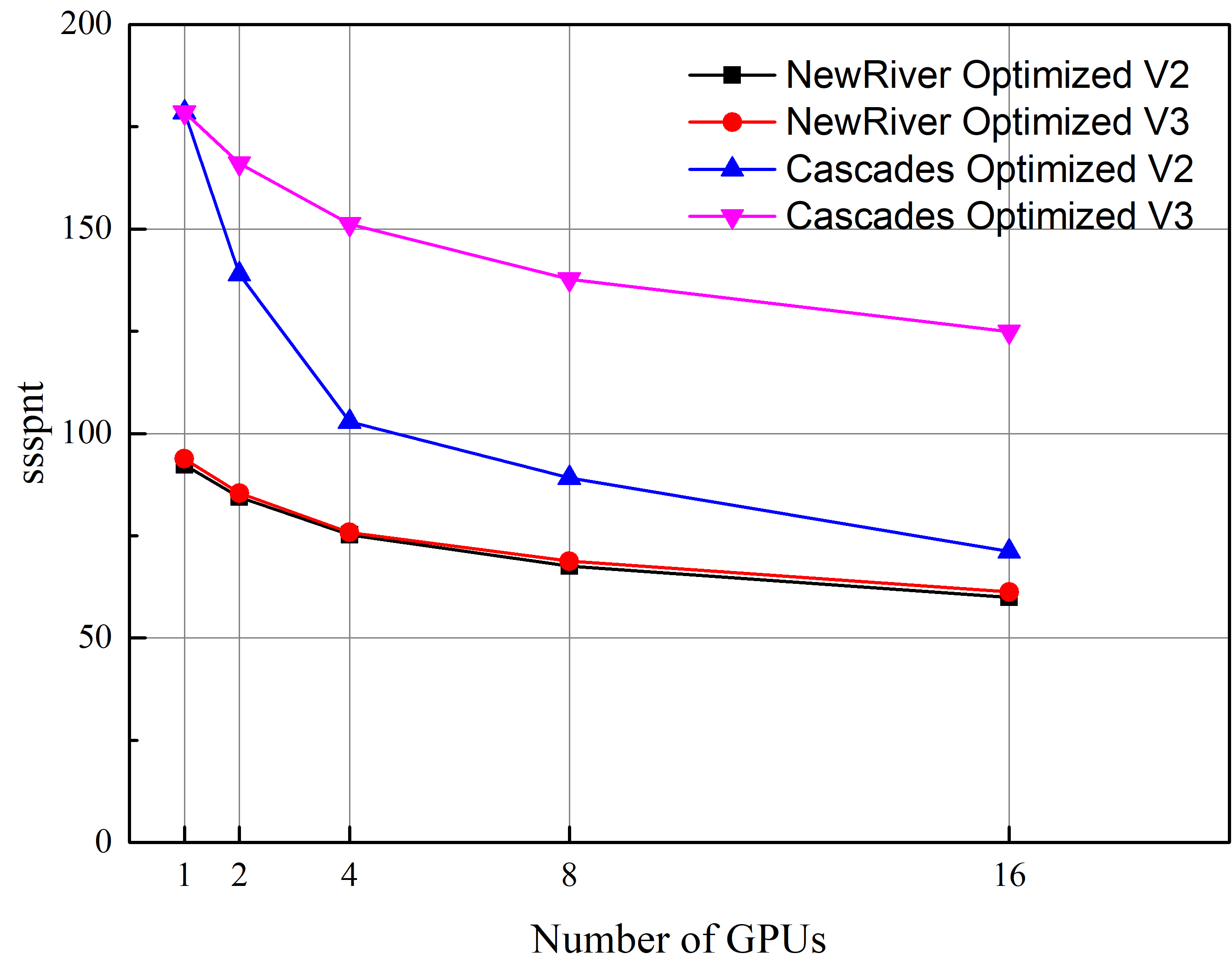} 
	}
	\subfigure[MVAPICH2]{ 
		\label{GPU_mvapich2_weak}
		\includegraphics[width=.30\textwidth]{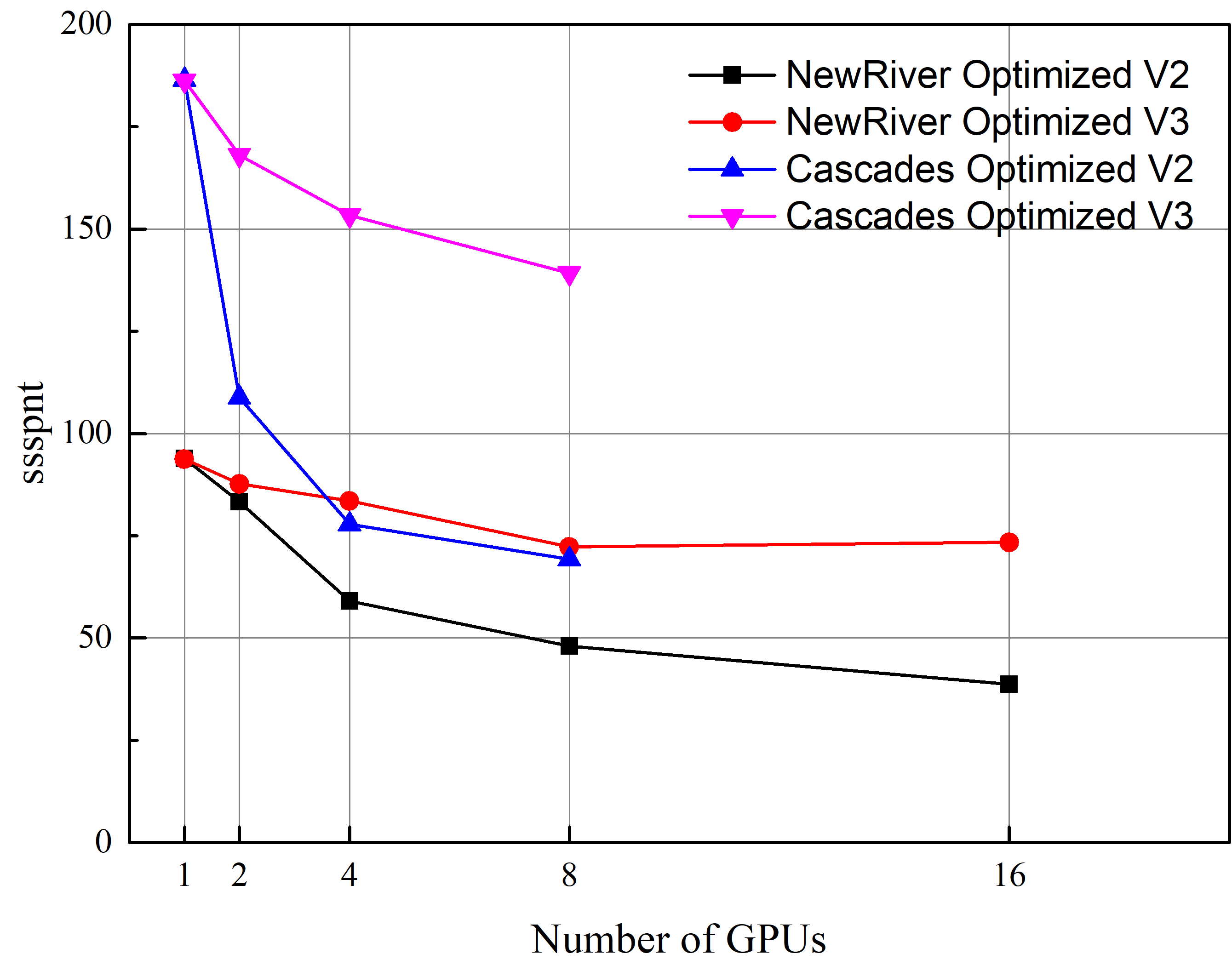}
	}
	\subfigure[MVAPICH2-GDR]{ 
		\label{GPU_mvapich2_gdr_weak}
		\includegraphics[width=.30\textwidth]{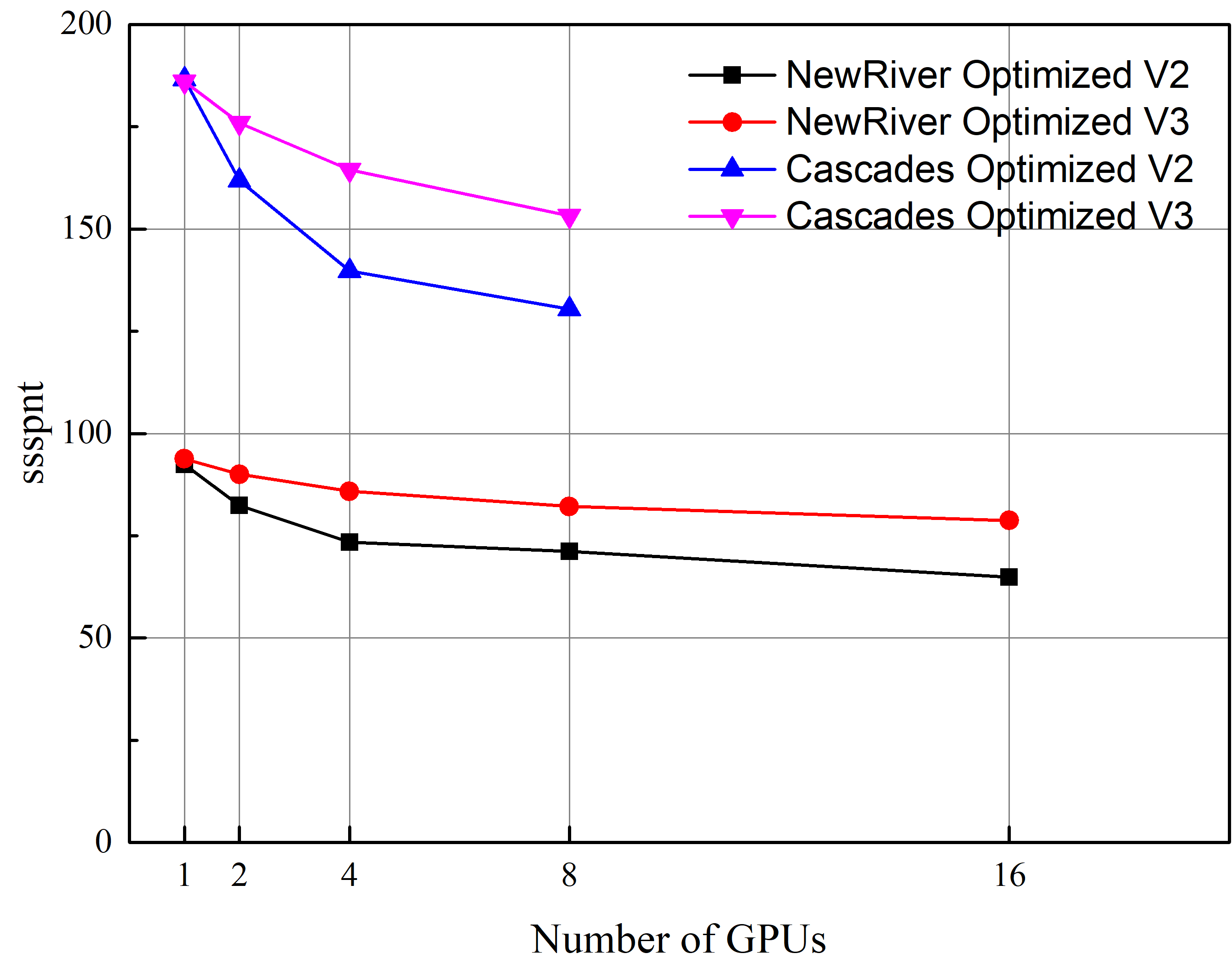}
	}	
	\caption{Weak scaling performance across platforms (3D decomposition)} 
	\label{GPUDirect_weak}
\end{figure}

When measuring the weak scaling performance, grid growth type 2 is applied. Fig.~\ref{GPUDirect_weak} shows the weak scaling performance using different MPI options. For each MPI option, results of using different decompositions for different grid growth are given. MVAPICH2 and Open MPI perform equivalently but there are still some differences. MVAPICH2 performs better than Open MPI for \emph{Optimized V3}, and generally worse for \emph{Optimized V2}, compared to Open MPI. It is reasonable as MVAPICH2 is designed to reduce communication overhead for complicated communication patterns, but there is some overhead associated with this optimization. It can also be seen that GPUDirect (with MVAPICH2-GDR) brings some performance benefits and performs the best for both \emph{Optimized V2} and \emph{Optimized V3}.

\subsection{Overlapping Communication and Computation}
When overlapping communication and computation, every decomposed block is further separated into two components: internal and external domains. For large enough problems, the internal domain will have significantly more grid points than the external domain. These internal points do not need data from other blocks so they can compute their updates while the communication is occurring for the external portion of the block. After communication is finished, the external domain continues to finish the remaining computation. Overlapping will not reduce latency but it can hide the latency caused by inter-block communication. In this paper, overlapping communication and computation was applied to both CPUs and GPUs. Communication is always done on CPUs while computation can be performed on CPUs or GPUs. 

\begin{figure}[H]
	\centering 
	\subfigure[CPU scaling]{ 
		\label{CPU_Overlap}
		\includegraphics[width=.45\textwidth]{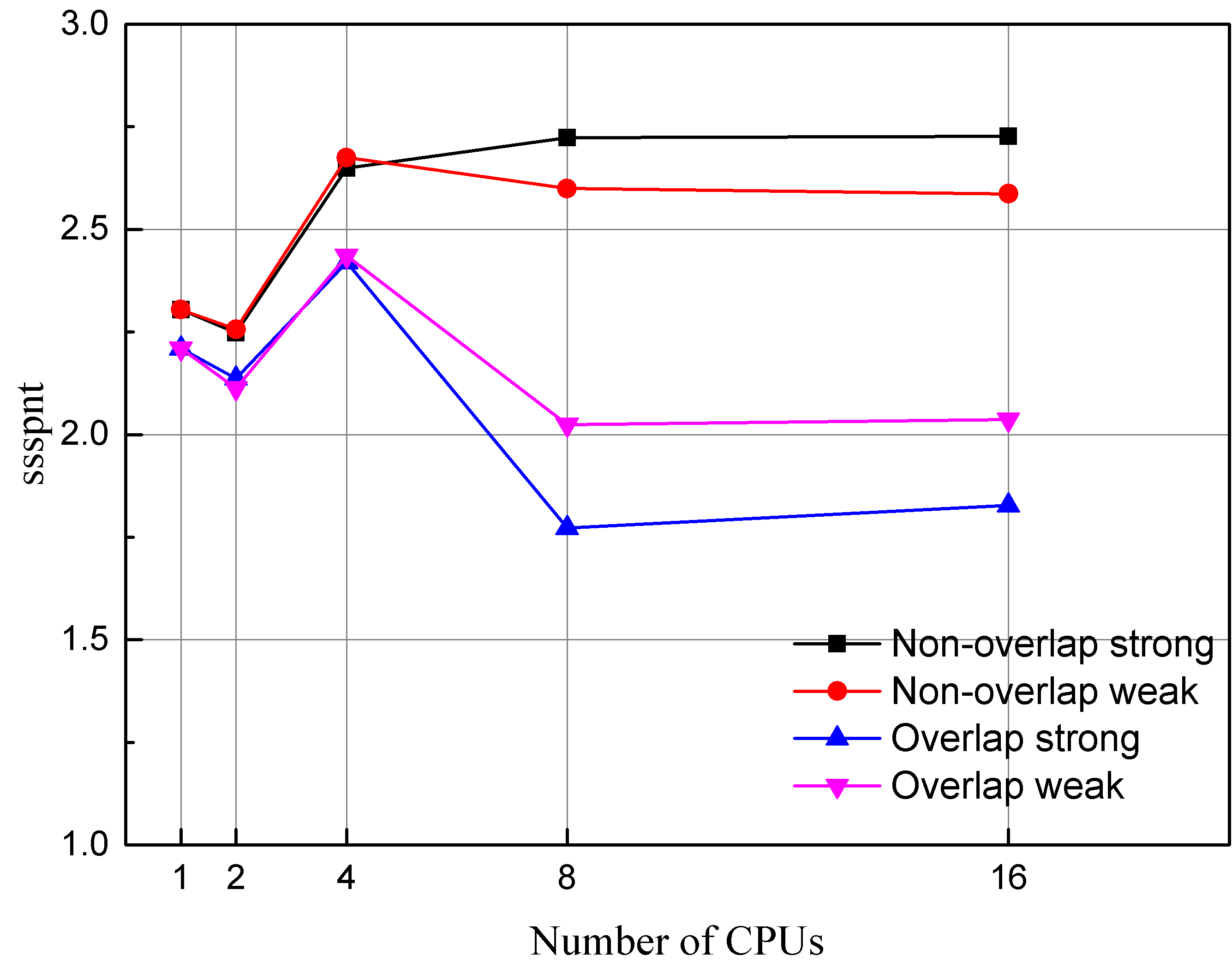} 
	} 
	\subfigure[GPU scaling]{ 
		\label{GPU_Overlap}
		\includegraphics[width=.45\textwidth]{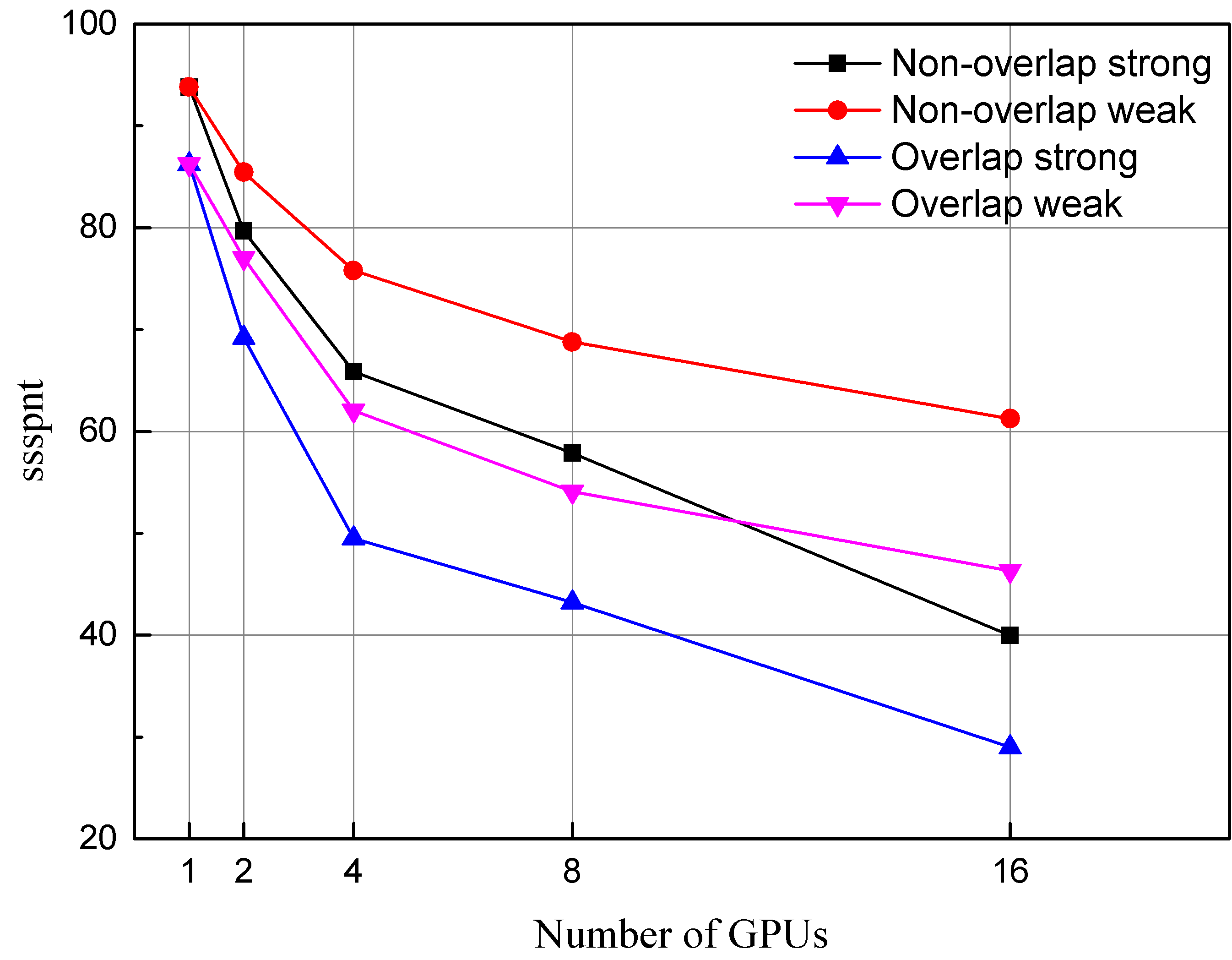}
	} 
	\caption{Overlap of communication and computation on NewRiver (3D decomposition)} 
	\label{Overlap}
\end{figure}

Case studies to compare the overlap and non-overlap versions have been made on different platforms, using different decompositions and code versions, and for the strong and weak scaling performance. The overlap version performs more slowly compared to the non-overlap version. Fig.~\ref{Overlap} shows the strong and weak scaling performance for both the CPU and GPU on the NewRiver cluster, with a performance comparison between the overlap version (extended from \emph{Optimized V3}) and the non-overlap (\emph{Optimized V3}) version. For both the CPU and GPU, overlap performs about 20\% to 30\% slower than non-overlap up to 16 processors, which was out of our initial expectation. The main reason is that the asynchronous progression is not supported well, potentially caused by the MPI and the communication system used. To figure out whether the asynchronous progression engine was activated or not, we used NVIDIA Visual Profiler~\cite{Profiler} to trace the program kernel executions on GPUs and found that the MPI used does not trigger communication until the code runs to a MPI\_WAITALL call, although communication is launched as early as possible. Since there is no actual overlap, and the non-overlap version only needs to setup the residual calculation kernel once while the overlap version has to do the setup multiple times (as it contains the internal domain and external domains), this overhead makes the overlap slower than the non-overlap version.

In fact, the MPI standard~\cite{MPI_standard} does not guarantee there is an actual overlap, which also means that the it may or may not be possible for communication to make progress when control has returned to the application, depending on the communication software and the underlying hardware. In Ref~\cite{bernholdt2008characterizing}, it is also concluded that the degree of actual overlap for an application depends on the overlap potential of both the application and the underlying communication subsystem. In our case, we tested the overlap on different Virginia Tech supercomputing platforms using different MPI options and different decompositions, and none of them improved the performance using multiple GPUs. It should be mentioned that the MPI standard allows for non-blocking operations to only be progressed to completion if a proper test/wait call was made. Thus, we tried to add many MPI{\_}TESTALL (dips into the MPI progression engine many times) or similar calls for the GPU code right after communication initialization. This makes overlap slightly better (observed through tracing). However, some overhead is produced due to adding these wait calls, also the degree of overlap is still not fully complete. The benefits of the performance enhancements are negligible compared with the overhead for our BDC code.

Our conclusion is that overlapping communication and computation is not a universal performance improvement for all applications and platforms, including the BDC problem using only MPI+OpenACC. Only if both the MPI compiler and the architecture supports asynchronous progression can overlap perform well and be used to hide some latency, which is difficult. An alternative way of improving the overlap is using MPI+OpenACC+OpenMP, in which OpenMP is used to generate multiple threads. These threads can work on different tasks such as computation and communication so that the actual degree of overlap can be increased \cite{jiayin2006overlapping,vaidyanathan2015improving,lu2015mpi+,denis2016mpi,castillo2019optimizing}. In fact, there are more literature discussing how to improve the overlap performance and almost all of them use multiple threads. Therefore, developers who have an interest in the overlap version for their own codes may need to do some simple tests first and should not only depend on overlap to get high performance. Since multi-threading is not a focus in this paper, no in-depth investigation of multi-threading is applied in this paper.

\section{Conclusions}

It is shown in the paper that OpenACC directives offer a convenient way to accelerate a CFD code fast on multiple platforms. All the platforms can generally use the same code with little code intrusion, which is a big advantage over CUDA and OpenCL. Some general optimizations are examined to improve the multi-GPU code performance, such as the pack/unpack method and stencil-based communication method. The optimizations introduced are shown to be very effective for both strong scaling and weak scaling, greatly reducing communication overhead and latency cost on GPUs. Further optimizations such as the overlap of communication and computation, asynchronous progression, and the use of CUDA-aware MPI and GPUDirect are also implemented and discussed. Overlapping communication and computation using only MPI+OpenACC is shown to be not an efficient way to improve the multi-GPU performance. GPUDirect is shown to be effective in a CFD application like the BDC code in this paper, as GPUDirect enables GPUs to communicate with each other directly and also increases the bandwidth between host and device. This avoids overhead between host and device and is important for communication-bound problems. Also, a combination of the use of GPUDirect and the optimizations proposed in this paper can improve both the strong and weak scaling performance substantially. 3D domain decomposition generally performs the best for the strong scaling on different platforms. For weak scaling, which decomposition performs best depends on how the grid growth is.

\section*{Acknowledgements}
The authors would like to thank Andrew J. McCall and Behzad Baghapour for creating the original BDC code as well as giving advice, and thank Charles W. Jackson for reviewing the paper and participating in various helpful discussions.

\bibliography{BDC_GPU_article}

\end{document}